\documentclass[lettersize,journal]{IEEEtran}
\usepackage{amsmath,amsfonts}
\usepackage{algorithmic}
\usepackage{algorithm}
\usepackage{array}
\usepackage[caption=false,font=normalsize,labelfont=sf,textfont=sf]{subfig}
\usepackage{textcomp}
\usepackage{stfloats}
\usepackage{url}
\usepackage{verbatim}
\usepackage{graphicx}
\usepackage{cite}
\usepackage{listings}
\usepackage{amsmath,amsfonts}
\usepackage{graphicx}
\usepackage{amsfonts}
\usepackage{array}
\usepackage{amssymb}
\usepackage{helvet}
\usepackage{color}
\usepackage{bm}
\usepackage{float}
\usepackage{cite}
\usepackage{multirow}
\usepackage{slashbox}
\usepackage{amssymb}
\usepackage[caption=false,font=normalsize,labelfont=sf,textfont=sf]{subfig}
\usepackage{amsmath,amsfonts}
\usepackage{algorithmic}
\usepackage{algorithm}
\usepackage{textcomp}
\usepackage{stfloats}
\usepackage{url}
\usepackage{verbatim}
\usepackage{graphicx}
\usepackage{cite}
\usepackage{amssymb}
\usepackage{makecell}
\usepackage{mathrsfs}

\newcommand{\RNum}[1]{\uppercase\expandafter{\romannumeral #1\relax}}

\newcounter{MYtempeqncnt}

\begin{document}

\title{6D Radar Sensing and Tracking in Monostatic
	 Integrated Sensing and Communications  System}
\author{Hongliang Luo, Feifei Gao, Fan Liu and Shi Jin
\thanks{H. Luo and F. Gao are with Department of Automation, Tsinghua University, Beijing 100084, China (email: luohl23@mails.tsinghua.edu.cn; feifeigao@ieee.org).}
\thanks{F. Liu is with the Department of Electrical and Electronic Engineering,
Southern University of Science and Technology, Shenzhen 518055, China
(e-mail: liuf6@sustech.edu.cn).}
\thanks{S. Jin is with the National Mobile Communications Research Laboratory, Southeast University, Nanjing 210096, China (e-mail:
jinshi@seu.edu.cn).
}
}



\maketitle

\begin{abstract}
In this paper, 
we propose a novel scheme for six-dimensional (6D) radar sensing and tracking of dynamic target based on multiple input and  multiple output (MIMO) array for monostatic integrated sensing and communications (ISAC) system.
Unlike most existing ISAC studies  believing that only the radial velocity of far-field dynamic target can be measured based on one single base station (BS), 
we find that the sensing echo channel of MIMO-ISAC system  actually includes the distance, horizontal angle, pitch angle, radial velocity, horizontal angular velocity, and pitch angular velocity of the dynamic target. Thus we may fully rely on one single BS to estimate the dynamic target's 6D motion parameters from the sensing echo signals.
Specifically, we first propose the  \emph{long-term motion} and \emph{short-term motion} model of dynamic target, in which the short-term motion model serves the \emph{single-shot sensing} of dynamic target, while the long-term motion model serves \emph{multiple-shots tracking} of dynamic target.
As a step further, we derive the   sensing channel model corresponding to the short-term motion. 
Next, for single-shot sensing, we employ the  array signal processing methods to estimate the dynamic target's  horizontal angle, pitch angle, distance, and \emph{virtual velocity}. By realizing that the virtual velocities observed by different antennas are different, we adopt plane  fitting to estimate the radial velocity, horizontal angular velocity, and pitch angular velocity of  dynamic target.
Furthermore, we implement the multiple-shots tracking of dynamic target  based on  each single-shot sensing results and Kalman filtering.
Simulation results  demonstrate the effectiveness of the proposed 6D radar sensing and tracking scheme.
\end{abstract}


\section{Introduction}

In the past decade, the integration of wireless communications and radar sensing has promoted the researches on
 dual functions radar communications (DFRC) systems\cite{8828023,8999605}. 
With the further expansion of the connotation and extension of sensing, integrated sensing and communications (ISAC) that incorporates more diverse sensing  technologies based on DFRC has been recognized as a promising air-interface technology for next-generation wireless networks\cite{202310141,9040264,9606831}.
Since ISAC  allows sensing systems and communications systems to share
  spectrum resources, and can serve various  intelligent applications, it has also been officially approved  by ITU-R IMT 2030 as one of the six key usage scenarios for the sixth generation (6G) mobile communications\cite{a11221,9755276}.

The ultimate functionality  of sensing is to construct the mapping relationship from  real physical world to  digital twin world,
where the former includes 
static environment (such as roads and buildings) and dynamic targets (such as pedestrians and vehicles). 
Therefore, realizing static environment reconstruction  and dynamic target sensing   is becoming  one consensus among researchers.
Specifically, 
dynamic target sensing,  as a research focus,   refers to the discovery, detection, parameters estimation, tracking, and recognition of  target based on the radar sensing function of  ISAC system.

On the other side, 
the base stations (BSs) in ISAC systems are usually stationary and are configured with massive multiple input and  multiple output (MIMO) arrays.
Depending on the number and location of BSs,  the ISAC system can be divided into: 
1) monostatic ISAC system (only one BS in the system); 
2) bistatic ISAC system (two BSs in the system); 
and 
3) multistatic ISAC system (multiple BSs in the system)\cite{9906898,2023arXiv230512994H}.
Among different ISAC architectures, monostatic ISAC system has received tremendous research attention due to low implementation complexity, as it does not require high-precision synchronization among BSs.

More relevant to this work, estimating the motion parameters of dynamic target with monostatic ISAC system has been well-investigated in the past few years.
For example,  X.~Chen~\emph{et.~al.}  proposed a multiple signal classification  based monostatic ISAC system that can  attain high  accuracy for target's angle, distance, and radial velocity estimation\cite{10048770}.
W.~Jiang~\emph{et.~al.}  proposed a model-driven   ISAC scheme, which simultaneously accomplished tasks of  demodulating  uplink communications signals and estimating  distance and radial velocity of dynamic target\cite{2023arXiv230715074J}.
In order to track  dynamic target,
F.~Liu~\emph{et.~al.} investigated a  sensing assisted predictive beamforming design for vehicle   communications  by exploiting  ISAC technique\cite{9171304}.
On top of that, 
Z.~Du~\emph{et.~al.}  proposed a   tracking scheme for extended  target based on extended Kalman filtering and  beamwidth adjustment, which  leveraged matched filtering and maximum likelihood estimation to obtain the angle, distance, and radial velocity of   target, and then tracks the target\cite{9947033}.
It should be noted that 
all of these works adopt the traditional view in the field of radar sensing that only the radial velocity of dynamic target can be measured based on one single BS, while the angular velocity  cannot be         directly        measured. Even in the field of radar sensing, the most advanced researches currently available suggest that the monostatic MIMO radar  can and only can realize the 4D sensing of dynamic target, namely,  measuring the dynamic target's horizontal angle, pitch angle, distance, and radial velocity\cite{9429942,9913510}.
However, by re-examining  the relationship between the  motion parameters of dynamic target  and the sensing echo channel of MIMO  system, 
one may realize  that the sensing  channel  already encompasses
 the distance, horizontal angle, pitch angle, radial velocity, horizontal angular velocity, and pitch angular velocity of the dynamic target.
 As a consequence, it becomes possible to  estimate the dynamic target's 6D motion parameters from the  echo signals based on one single BS.

Evidently, there are already
 some preliminary studies focusing on measuring the angular velocity based on monostatic radar system. 
J.~A.~Nanzer~\emph{et.~al.}  first proposed the theoretical method for measuring the angular velocity of moving object based on spatial interferometry using one single station radar system, 
which was later verified through hardware experiments\cite{5634150,6697333,6711074,6236072}.
X.~Wang~\emph{et.~al.}  extended this work to multiple targets angular velocities  measurement scenarios through conceiving sophisticated algorithms\cite{8835771,8936370,9114597}.
However, all of these studies only considered the spatial interference effect between two or three antennas, resulting in severe sensing performance loss. 
To the best knowledge of the authors, estimating the  angular velocity based on single  station massive MIMO-OFDM system still remains widely unexplored.

In this paper, we attempt to fill in this research gap by proposing 
 a novel scheme for 6D radar single-shot sensing and multiple-shots  tracking of dynamic target based on massive MIMO array for monostatic ISAC system.
The contributions of this paper are summarized as follows.

\begin{itemize}
	
\item Based on  the working pipeline of  radar sensing  in  ISAC system, we  construct the long-term motion  and short-term motion model for dynamic target   in  3D space, 
which correspond to multiple-shots tracking and single-shot sensing of dynamic target,
respectively.

\item We re-examine  the relationship between the 6D motion parameters of dynamic target  and the sensing echo channel of  MIMO-ISAC system,
and reveal that the sensing  channel  actually includes the distance, horizontal angle, pitch angle, radial velocity, horizontal angular velocity, and pitch angular velocity of   target, which enables us to  estimate the dynamic target's 6D motion parameters from   echoes by solely relying on one single ISAC BS.

\item For single-shot sensing, we employ  the array signal processing methods to estimate the dynamic target's distance, horizontal angle, pitch angle, and virtual velocity. Then we show  that the virtual velocities observed at different antennas are
distinct from each other,  allowing us to utilize plane  fitting to estimate the radial velocity, horizontal angular velocity, and pitch angular velocity of the dynamic target.

\item Based on the single-shot 6D parameters sensing results, we further propose a multiple-shots tracking approach for dynamic target through Kalman filtering.

	
\end{itemize}

The remainder of this paper is organized as follows.
In Section \RNum{2}, we propose the 6D motion model of dynamic target, and derive the  corresponding sensing channel model.
In Section~\RNum{3}, we propose
a novel 6D sensing and tracking scheme for dynamic target sensing.
Simulation results and conclusions are given in Section~\RNum{4} and Section~\RNum{5},
respectively.

\emph{Notation}:
Lower-case and upper-case boldface letters $\mathbf{a}$ and $\mathbf{A}$ denote a vector and a matrix;
$\mathbf{a}^T$ and $\mathbf{a}^H$ denote the transpose and the conjugate transpose of vector $\mathbf{a}$, respectively;
$[\mathbf{a}]_n$  denotes the $n$-th element of the vector $\mathbf{a}$;
$[\mathbf{A}]_{i,j}$ denotes the $(i,j)$-th element of the matrix $\mathbf{A}$; $\mathbf{A}[i_1:i_2,:]$ is the submatrix composed of all columns elements in rows $i_1$ to $i_2$ of matrix $\mathbf{A}$;
$\mathbf{A}[:,j_1:j_2]$ is the submatrix composed of all rows elements in columns $j_1$ to $j_2$ of matrix $\mathbf{A}$;
${\rm eig}(\cdot)$ represents the matrix eigenvalue decomposition function.

\section{System Model and Proposed ISAC Framework}

In this section, we provide the generic model for massive MIMO based monostatic ISAC system, and   propose the 6D motion model of dynamic target, as well as derive the   sensing channel model of  ISAC system. 

\vspace{-4mm}

\subsection{ISAC BS Model}

\begin{figure}[!t]
	\centering
	\includegraphics[width=80mm]{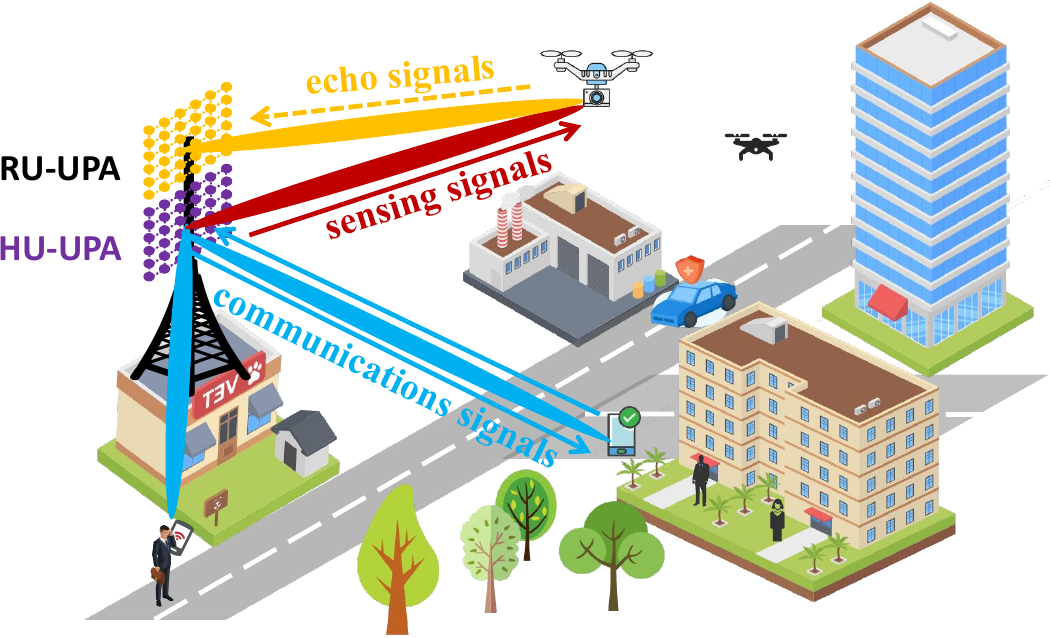}
	\caption{System model.}
	\label{fig_1}
\end{figure}

A massive MIMO based monostatic ISAC system
operating in  mmWave or  Terahertz frequency bands with OFDM modulation is depicted in Fig.~1, which employs only one dual-functional BS for wireless communications and radar sensing at the same time.
Generally, we consider that the BS consists of one \emph{hybrid unit (HU)} and one \emph{radar  unit (RU)}, where both the HU and the RU
are configured with uniform planar arrays (UPAs). 
By designing the beamforming strategy, HU is responsible for transmitting downlink communications signals and receiving uplink communications signals, as well as transmitting downlink sensing  signals to realize dynamic target sensing; while 
RU is   responsible for receiving   echo signals to realize dynamic target  sensing.

The HU and  RU are each equipped with one UPA  of $N_H=N^{x}_{H}\times N_{H}^z$ and $N_R=N_{R}^x\times N_{R}^z$ antenna elements,  named as
HU-UPA and RU-UPA, respectively. 
Assume that both the HU-UPA and the  RU-UPA are   vertically mounted on the 2D plane $y = 0$ at BS side as shown in Fig.~2,
and the antenna spacing between the antennas distributed along  x-axis and z-axis are 
$d_x = d\le \frac{\lambda}{2}$ and 
$d_z = d\le \frac{\lambda}{2}$, respectively,
with $\lambda$ being the wavelength. 
Without loss of generality, we denote the
position of the $n_H$-th antenna element  in  the HU-UPA as $\mathbf{p}_{n_H}=\mathbf{p}_{0_H}+[d\cdot n_{H}^x,0,d\cdot n_{H}^z]^T$, where
$\mathbf{p}_{0_H}$ is the position of the reference  element, 
 $n_{H}^x \in \{0,1,...,N_{H}^x-1\}$ and $n_{H}^z \in \{0,1,...,N_{H}^z-1\}$ are the antenna indices. 
Here we  use two types of index numbers to represent the same antenna,
that is, the $n_H$-th antenna may also be named as the $(n^x_H,n^z_H)$-th antenna.
Similarly, we denote the
position of the $n_R$-th antenna element  in  the RU-UPA as $\mathbf{p}_{n_R}= \mathbf{p}_{0_R} + [d\cdot n_R^x,0,d\cdot n_R^z]^T$ with
$n_{R}^x \in \{0,1,...,N^x_{R}-1\}$ and $n_{R}^z  \in \{0,1,...,N^z_{R}-1\}$.
Generally, according to the spatial consistency of the arrays, 
we further  assume that  the HU-UPA and  RU-UPA
are co-located and are parallel to each other, i.e., $\mathbf{p}_{0_H} = \mathbf{p}_{0_R} = [0,0,0]^T$, such that they may see the targets at the same propagation directions\footnote{Since the normal communications and sensing distance is much longer than the protection distance between  HU-UPA and RU-UPA, it can be considered that HU-UPA and RU-UPA are located in the same position.}\cite{9898900}.
Besides, by balancing the hardware costs and system performance,
we assume that the HU-UPA employs the  hardware architecture based on  phase shifter (PS) structure, in which a total of $N_{H,RF}\ll N_H$ radio frequency (RF) chains are deployed, and each antenna is connected to a PS  to realize beamforming.  
On the other side, we  assume that the  RU-UPA employs the fully-digital receiving array, with each antenna connected to one RF chain,  to realize super-resolution sensing  performance, which follows the setting in \cite{10048770}  and \cite{9898900}.

\begin{figure}[!t]
	\centering
	\includegraphics[width=65mm]{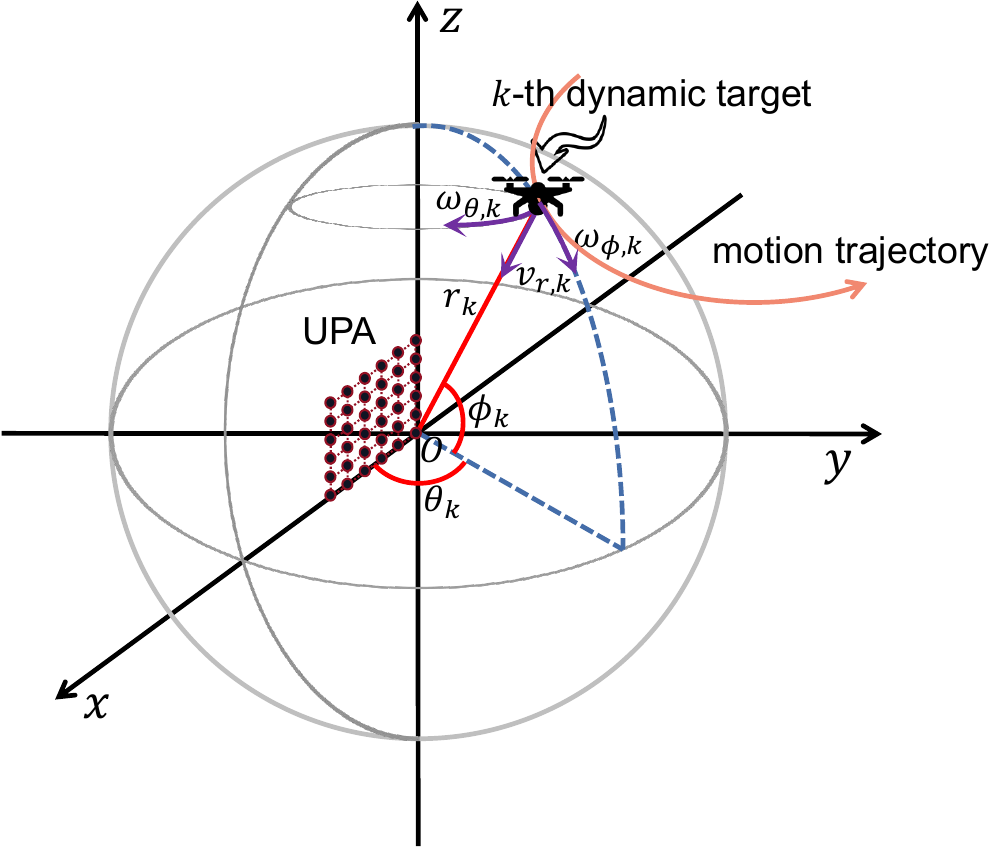}
	\caption{BS model and dynamic target motion parameters.}
	\label{fig_1}
\end{figure}

\begin{figure}[!t]
	\centering
	\includegraphics[width=90mm]{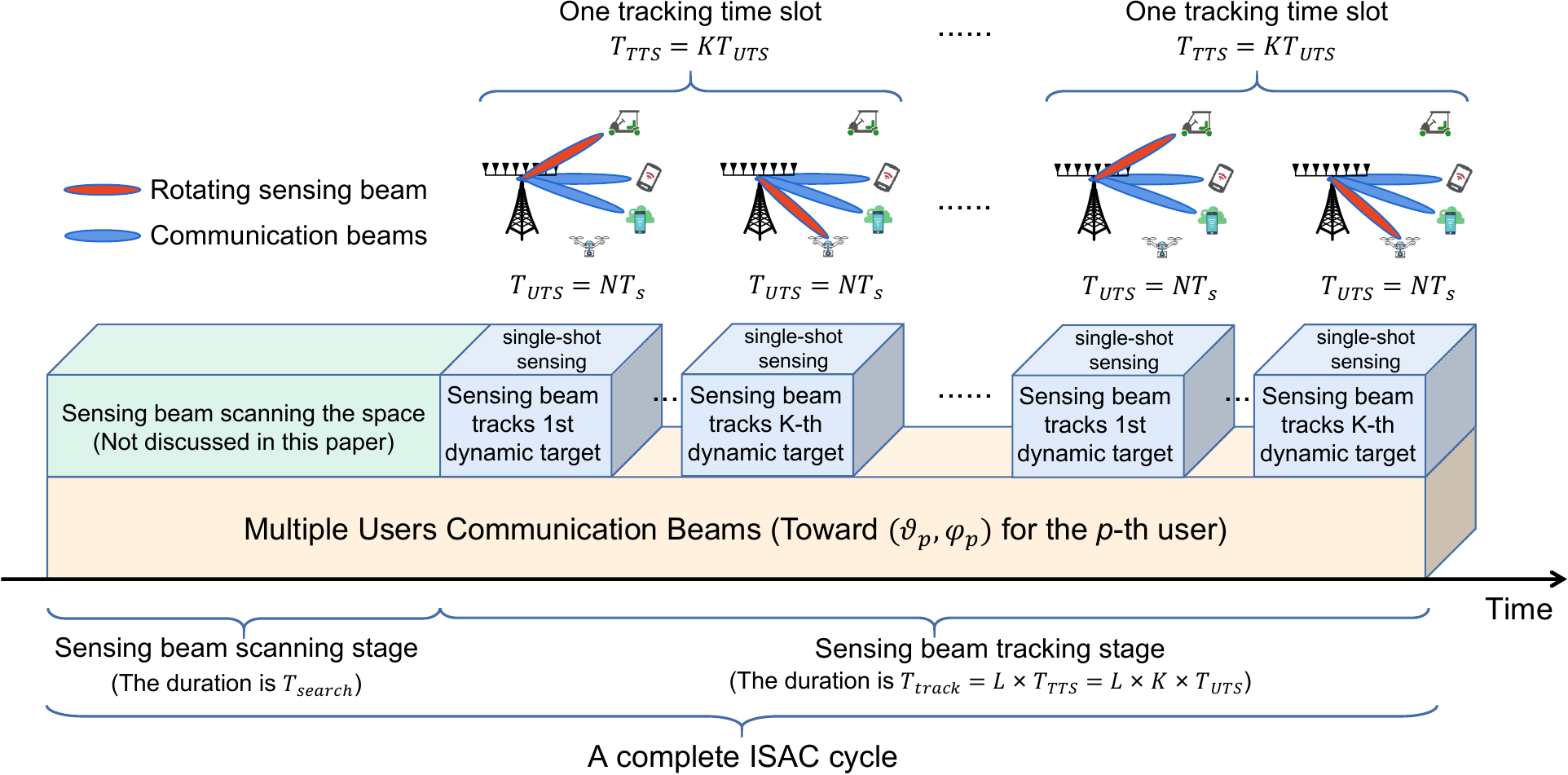}
	\caption{The proposed ISAC framework.}
	\label{fig_1}
\end{figure}

Suppose that the ISAC system emits   OFDM signals with $M$ subcarriers,
where the  lowest  frequency and the subcarrier  interval of OFDM signals are $f_0$ and $\Delta f$, respectively.
Then the transmission bandwidth is $W=(M-1)\Delta f$, and 
the frequency of the $m$-th subcarrier  is $f_m=f_0+m\Delta f$, where $m=0,1,...,M-1$. 
Further, we consider that an OFDM frame contains $N$ consecutive OFDM symbols, where  
the time interval between adjacent OFDM symbols is  $T_s = T'_s+T_g$, 
with $T'_s = \frac{1}{\Delta f}$ and $T_g$
being the OFDM symbol duration and guard interval, respectively.

We  employ the spherical coordinates $(r,\theta,\phi)$ to represent one  position in 3D space. As shown in Fig.~2, $r$  represents the polar distance with  mathematical range of $r \geq 0$,
$\theta$ represents  the horizontal angle with  mathematical range of $0^\circ \leq \theta \leq 180^\circ$, and $\phi$  represents the pitch angle with  mathematical range of $-90^\circ \leq \phi \leq 90^\circ$. 
Moreover, the spherical coordinate $(r,\theta,\phi)$ 
may be translated to its Cartesian counterpart  $(x,y,z)$ through 
\begin{align}
	x=r\cos \phi \cos \theta,
	\quad y=r\cos \phi \sin \theta,
	\quad z=r\sin \phi \label{3}.
\end{align}

Since  BS is located at the origin of  coordinate system, we  denote 
the service area of   BS as 
$\{(r,\theta,\phi)|r_{min}\leq r \leq r_{max},\theta_{min}\leq \theta \leq \theta_{max},\phi_{min}\leq \phi \leq \phi_{max}\}$. 
Suppose that there are $P$ single-antenna communications users, $K$ dynamic targets, as well as widely distributed static environment within this service area.
We assume that the 6D motion parameters of the $k$-th dynamic target are
$\{r_k, \theta_k, \phi_k, v_{r,k}, \omega_{\theta,k}, \omega_{\phi,k}\}$,
in which $(r_k,\theta_k,\phi_k)$ represents the position of the $k$-th target, 
$v_{r,k}$,  $\omega_{\theta,k}$ and $\omega_{\phi,k}$ represent the radial velocity, horizontal angular velocity, and pitch angular velocity, respectively.
Besides,
we  assume that  the position of the $p$-th user is $(R_p,\vartheta_p,\varphi_p)$,
which are known and  stationary to  BS, due to the fact that they can  be easily obtained through  user reporting,  or other techniques\cite{9782674,10271123,9013639}.

\vspace{-4mm}

\subsection{The Proposed ISAC Framework}

The task of  ISAC system is to sense all $K$ dynamic targets while serving the communications of all $P$ users.
As described in Fig.~3,
the proposed ISAC framework consists of two stages: \emph{sensing beam scanning (SBS) stage} and  \emph{sensing beam tracking (SBT) stage}.
For the aspect of communications,   BS continuously generates $P$ communications beams  towards $P$    users
to  maintain  communications service during both SBS stage and SBT stage. 
For the aspect of sensing,  BS 
generates one  sensing beam that can  scan 
the  service area during  SBS stage, 
during which the  BS may detect the  targets
and estimate their parameters.
To proceed,  BS generates a single sensing beam  to track all $K$  dynamic targets in a time division manner during  SBT stage,
that is, the sensing beam may sequentially illuminate each  target and continuously track them.
In this work, we  mainly focuses on the  SBT stage,
and refer readers to our previous work \cite{2023arXiv231101674L} for more details on the SBS stage.

Assume that $K$ dynamic targets possess different physical  directions.
At each direction,  the BS may adopt one OFDM frame, i.e., $N$ consecutive OFDM symbols,  to realize dynamic target sensing. 
As shown in Fig.~3, we  divide the SBT stage  into $L$ tracking time slots (TTSs),  
and each TTS lasts for a time duration of $T_{TTS} = K\cdot N\cdot T_s$.
Besides, each TTS is further divided into $K$ unit time slots (UTSs), and each UTS lasts for a time duration of $T_{UTS} =  N\cdot T_s$. Clearly, there is $T_{TTS} = K\cdot T_{UTS}$.
During  $L$ consecutive TTSs,  BS should  track all $K$ dynamic targets using the methods such as Kalman filtering with $T_{TTS}$ as  time step, which is named 
as \emph{multiple-shots tracking}.
To realize continuous target tracking and reliable users communications,
in the $(l,k')$-th UTS, BS needs to generate $P$ communications beams pointing to $P$ users and  generate one sensing beam pointing to the \emph{sensing tracking direction} $(\xi_{lk'},\eta_{lk'})$
from  HU-UPA, where $k'=1,2,...,K$. 
The BS needs to update the motion parameter sensing results of the $k'$-th dynamic target within this UTS,
known
as \emph{single-shot sensing}.
Ideally, $(\xi_{lk'},\eta_{lk'})$ should be equal to the physical  direction of the $k'$-th dynamic target in the $(l,k')$-th UTS.

Here we assume that the transmission power of  BS is $P_t$,
the energy of sensing beam is  $\rho_{lk'} P_t$,
and the remaining energy $(1-\rho_{lk'})P_t$ is evenly distributed among  communications beams, where $\rho_{lk'} \in [0,1]$ is the power distribution coefficient used in the $(l,k')$-th UTS.
Due to the short duration of one UTS, we keep the directions of   transmitting beams   unchanged within one UTS.
Then the transmission signals from  HU-UPA  on the $m$-th subcarrier of the $n$-th  symbol in the $(l,k')$-th UTS should be represented as
\begin{equation}
	\begin{split}
		\begin{aligned}
			\label{deqn_ex1a}
\!\!\!& \mathbf{x}_{lk'\!,n,m} = \sum_{p=1}^{P}\mathbf{w}_{c,p,lk'}s^{c,p,lk'}_{n,m}+\mathbf{w}_{s,lk'}s^{s,lk'}_{n,m}
\\& \!\!\!= \!\!\! \sum_{p=1}^{P}\!\!\!\sqrt{\!\!\frac{(\!1\!\!-\!\!\rho_{lk'})\!P_{\!t}}{PN_H}}\!\mathbf{a}_{\!H}\!(\Gamma_p\!,\!\!\Upsilon_p\!)\!s^{c,p,lk'}_{n,m}\!\!\!\!+\!\!\!\sqrt{\!\frac{\rho_{lk'}\!P_{\!t}}{N_H}}\!\mathbf{a}_{\!H}\!(\Xi_{lk'}\!,\!\Theta_{lk'}\!)\!s^{s,lk'}_{n,m}\!,
		\end{aligned}
	\end{split}
\end{equation}
where $(\Gamma_p, \Upsilon_p) \!=\! (\cos \varphi_p \cos \vartheta_p,\sin \varphi_p)$ is the \emph{spatial-domain direction} corresponding to the \emph{physical  direction} $(\vartheta_p,\varphi_p)$ of the $p$-th user,
and $(\Xi_{lk'},\Theta_{lk'})=(\cos \eta_{lk'} \cos \xi_{lk'},\sin \eta_{lk'})$ is the spatial-domain direction corresponding to the  sensing tracking direction  $(\xi_{lk'},\eta_{lk'})$.
Besides, 
 $\mathbf{w}_{c,p,lk'}=\sqrt{\frac{(1-\rho_{lk'})P_t}{PN_H}}\mathbf{a}_{H}(\Gamma_p, \Upsilon_p)$ and $\mathbf{w}_{s,lk'}=\sqrt{\frac{\rho_{lk'}P_t}{N_H}}\mathbf{a}_{H}(\Xi_{lk'},\Theta_{lk'})$ are the communications beamforming vector for the $p$-th user  and the sensing beamforming vector for the  sensing tracking direction, respectively, and  $\mathbf{a}_{H}(\Psi,\Omega)$ is the array   steering vector  of HU-UPA with the form
\begin{equation}
	\begin{split}
		\begin{aligned}
			\label{deqn_ex1a}
\mathbf{a}_{H}(\Psi,\Omega)\! =
\mathbf{a}_{H}^x(\Psi)\otimes \mathbf{a}_{H}^z(\Omega)
	 \in\! \mathbb{C}^{N_H\times 1},
		\end{aligned}
	\end{split}
\end{equation}
where $\otimes$ denotes the Kronecker product, and 
\begin{align}
\!\!\mathbf{a}_{H}^x(\Psi)&\!=\![1,e^{j\frac{2\pi f_0d\Psi}{c}},...,e^{j\frac{2\pi f_0d\Psi}{c}(N_{H}^x-1)}]^T \!\!\in\! \mathbb{C}^{N_{H}^x\times 1} \label{1},\\
\!\!\mathbf{a}_{H}^z(\Omega)&\!=\! [1,e^{j\frac{2\pi f_0d\Omega}{c}},...,e^{j\frac{2\pi f_0d\Omega}{c}(N_{H}^z-1)}]^T \!\! \in\! \mathbb{C}^{N_{H}^z\times 1} \label{2}.
\end{align}
Moreover,  $s^{c,p,lk'}_{n,m}$ and $s^{s,lk'}_{n,m}$ are  communications signals for the $p$-th user and  
sensing detection signals, respectively.

Based on (2), the BS may realize both communications function and sensing function by optimizing $\rho_{lk'}$ during each  UTS, 
which may be conceived through the power allocation strategy proposed in 
 \cite{2023arXiv231101674L}, via maximizing the sensing performance while ensuring users communications performance. 
To that end,   we only consider   dynamic target sensing    problem while omitting the design of communications function in this work. 
Consequently, there is always one beam  towards the sensing  tracking direction $(\xi_{lk'},\eta_{lk'})$, and  we can rewrite the transmission signals from the HU-UPA  on the $m$-th subcarrier of the $n$-th OFDM symbol in the $(l,k')$-th UTS as
\begin{equation}
	\begin{split}
		\begin{aligned}
			\label{deqn_ex1a}
\mathbf{x}_{lk',n,m} =  \sqrt{\frac{\grave{\rho}_{lk'}P_t}{N_H}}\mathbf{a}_{H}(\Xi_{lk'},\Theta_{lk'})s^{t,lk'}_{n,m},
		\end{aligned}
	\end{split}
\end{equation}
where $\grave{\rho}_{lk'}P_t$ is  the power allocated to  $(\xi_{lk'},\eta_{lk'})$ direction, and $s^{t,lk'}_{n,m}$ is the  signal transmitted to  $(\xi_{lk'},\eta_{lk'})$ direction.

\vspace{-4mm}

\subsection{The 6D Motion Model of Dynamic Target}

The motion of dynamic target can be described from two levels: 1) long-term motion, and 2) short-term motion. 
Specifically, the dynamic target undergoes long-term motion within $L$ TTSs. Given the long tracking time of the target, the radial velocity and  angular velocities of the target are susceptible to various disturbances. In long-term motion, the time interval for describing the target motion is $T_{TTS}$. 
We refer to the 6D motion parameters of the $k$-th target at the $l$-th TTS as the state of this target, denoted as
\begin{equation}
	\begin{split}
		\begin{aligned}
			\label{deqn_ex1a}
\mathbf{S}^{Long}_{k,l}\!=\![r^{L\!o\!n\!g}_{k,l}, \theta^{L\!o\!n\!g}_{k,l}, \phi^{L\!o\!n\!g}_{k,l}, v^{L\!o\!n\!g}_{r,k,l}, \omega^{L\!o\!n\!g}_{\theta,k,l}, \omega^{L\!o\!n\!g}_{\phi,k,l}]^T \!\in\! \mathbb{R}^{6\times 1},
		\end{aligned}
	\end{split}
\end{equation} 
 where $l=0,1,...,L-1$. 
Without loss of  generality,  during the  motion process, the direction in which the polar distance decreases, the direction in which the horizontal angle value decreases, and the direction in which the pitch angle value decreases  are taken as the positive directions for radial velocity, horizontal angular velocity, and pitch angular velocity, respectively.
Then the 6D motion model of the $k$-th dynamic target for long-term motion can be represented as
\begin{align}
	r^{Long}_{k,l+1}&= r^{Long}_{k,l} - v^{Long}_{r,k,l}T_{TTS} - \frac{1}{2} u^{Long}_{r,k,l} T_{TTS}^2  \label{1},\\
	\theta^{Long}_{k,l+1}&=\theta^{Long}_{k,l}-\omega^{Long}_{\theta,k,l}T_{TTS}-\frac{1}{2} u^{Long}_{\theta,k,l} T_{TTS}^2
	 \label{2},\\
	\phi^{Long}_{k,l+1}&=\phi^{Long}_{k,l} - \omega^{Long}_{\phi,k,l}T_{TTS} -
	\frac{1}{2} u^{Long}_{\phi,k,l} T_{TTS}^2
	\label{3},\\
	v^{Long}_{r,k,l+1}&=v^{Long}_{r,k,l} - u^{Long}_{r,k,l} T_{TTS}
	\label{4},\\
	\omega^{Long}_{\theta,k,l+1}&=\omega^{Long}_{\theta,k,l}- u^{Long}_{\theta,k,l} T_{TTS}
	 \label{5},\\
	\omega^{Long}_{\phi,k,l+1}&=\omega^{Long}_{\phi,k,l}-
	 u^{Long}_{\phi,k,l} T_{TTS}, \label{6}
\end{align}
where $\mathbf{u}^{Long}_{k,l}=[u^{Long}_{r,k,l},u^{Long}_{\theta,k,l},u^{Long}_{\phi,k,l}]^T$
represents the random disturbance during the long-term   motion.

In addition to sense the long-term motion, the BS needs to observe the $k'$-th dynamic target within the $(l,k')$-th UTS. 
The dynamic target  also undergoes short-term motion within this UTS. 
Due to the short duration of the short-term  motion, 
one usually assumes that the velocities of the dynamic target remain constant within one UTS time, i.e., $N$ OFDM symbol times\cite{10048770}.
Then the 6D motion parameters of the $k$-th dynamic target within the $n$-th OFDM symbol time of the $(l,k')$-th UTS can be expressed as 
\begin{equation}
	\begin{split}
		\begin{aligned}
			\label{deqn_ex1a}
\!\!\!\!\!\!\!\! \mathbf{S}^{Short}_{k\!,lk'\!\!,n}\!\!=\!\![r^{Short}_{k\!,lk'\!\!,n},\! \theta^{Short}_{k\!,lk'\!\!,n},\! \phi^{Short}_{k\!,lk'\!\!,n},\! v^{Short}_{r,k\!,lk'\!\!,n},\! \omega^{Short}_{\theta,k\!,lk'\!\!,n},\! \omega^{Short}_{\phi,k\!,lk'\!\!,n}]^T 
		\end{aligned}
	\end{split}
\end{equation}  
 with $n=0,1,...,N-1$, which satisfies 
\begin{align}
	r^{Short}_{k,lk',n}&= r^{Long}_{k,l} - v^{Long}_{r,k,l}nT_{s}   \label{1},\\
	\theta^{Short}_{k,lk',n}&=\theta^{Long}_{k,l}-\omega^{Long}_{\theta,k,l}nT_{s}
	\label{2},\\
	\phi^{Short}_{k,lk',n}&=\phi^{Long}_{k,l} - \omega^{Long}_{\phi,k,l}nT_{s} 
	\label{3},\\
	v^{Short}_{r,k,lk',n}&=v^{Long}_{r,k,l} 
	\label{4},\\
	\omega^{Short}_{\theta,k,lk',n}&=\omega^{Long}_{\theta,k,l}
	\label{5},\\
	\omega^{Short}_{\phi,k,lk',n}&=\omega^{Long}_{\phi,k,l}. \label{6}
\end{align}

Naturally, 
the long-term motion model corresponds to the multiple-shots tracking of dynamic target, while short-term motion model corresponds to single-shot sensing of dynamic target.
The ISAC system needs to utilize    $\mathbf{S}^{Short}_{k,lk',n}$ as much as possible to sense and track  $\mathbf{S}^{Long}_{k,l}$ of the $k$-th dynamic target, that is, the ISAC system needs to utilize single-shot sensing to realize  multiple-shots tracking.

\vspace{-5mm}

\subsection{Sensing Channel Model of ISAC System}

In the $(l,k')$-th UTS, the BS transmits   the   detection signals through HU-UPA at the beginning of sensing, which will be reflected by dynamic targets and cause echoes. Then, the RU-UPA   will receive the sensing echo signals.
Let us define  the path from the $n_H$-th antenna of  HU-UPA to the $k$-th dynamic target and then back to the $n_R$-th antenna of RU-UPA as the $(n_H,k,n_R)$-th propagation path. 
Then we denote 
 $\tau^{lk'\!,n}_{k,n_H\!,n_R}\!=\!(D^{lk',n}_{k,n_H}+D^{lk',n}_{k,n_R})/c$ 
as the time delay of the $(n_H,k,n_R)$-th propagation path in the $n$-th OFDM symbol time during the $(l,k')$-th UTS,
where $c$ represents the speed of light, $D^{lk',n}_{k,n_H}$ is the distance between the $n_H$-th antenna of HU-UPA and the $k$-th dynamic target, and $D^{lk',n}_{k,n_R}$ is the distance between the $n_R$-th antenna of RU-UPA and the $k$-th dynamic target.

Suppose that the signal transmitted by the
$n_H$-th antenna of  HU-UPA is $s(t)$,
 and the corresponding passband signal is  $\mathcal{R}\{s(t)e^{j2\pi f_0t}\}$. 
Then the echo signal will be a delayed version of the transmitting signal with amplitude attenuation.
Specifically, the passband echo signal received by the $n_R$-th antenna
through the $(n_H,k,n_R)$-th propagation path
at the $n$-th OFDM symbol during  the $(l,k')$-th UTS  is
$\mathcal{R}\{\alpha^{lk'}_{k} s(t-\tau^{lk',n}_{k,n_H,n_R})e^{j2\pi f_0(t-\tau^{lk',n}_{k,n_H,n_R})}\}$.
The corresponding baseband echo signal is
$\alpha^{lk'}_{k} s(t-\tau^{lk',n}_{k,n_H,n_R})e^{-j2\pi f_0\tau^{lk',n}_{k,n_H,n_R}}$, 
and  the baseband  equivalent channel is
\begin{equation}
\begin{split}
\begin{aligned}
\label{deqn_ex1a}
h^{lk',n}_{k,n_H,n_R}(t) = \alpha^{lk'}_k e^{-j2\pi f_0\tau^{lk',n}_{k,n_H,n_R}} \delta \left(t-\tau^{lk',n}_{k,n_H,n_R}\right),
\end{aligned}
\end{split}
\end{equation}
where $\alpha^{lk'}_k$ is the  channel fading factor and $\delta (\cdot)$ denotes the Dirac delta function.  Taking the Fourier transform of (21), the baseband frequency-domain channel response can be obtain as
\begin{equation}
	\begin{split}
		\begin{aligned}
		\label{deqn_ex1a}
h^{lk',n,F}_{k,n_H,n_R}(f) = \alpha^{lk'}_k e^{-j2\pi (f_0+f)\tau^{lk',n}_{k,n_H,n_R}}.
		\end{aligned}
	\end{split}
\end{equation}
Thus the frequency-domain sensing echo channel  of the $(n_H,k,n_R)$-th propagation path    on the $m$-th subcarrier of the $n$-th OFDM symbol during the $(l,k')$-th UTS is 
\begin{equation}
	\begin{split}
		\begin{aligned}
			\label{deqn_ex1a}
\!\!\!\!\!\!\!\! h^{lk',n,m}_{k,n_H,n_R} \!\!=\! \alpha^{lk'}_k  \!\!  e^{-j2\pi f_m\tau^{lk',n}_{k,n_H,n_R}}  \!=\!\alpha^{lk'}_k \!\! e^{\!-j2\pi f_m \frac{D^{lk',n}_{k,n_H}\!+D^{lk',n}_{k,n_R}}{c}}.
		\end{aligned}
	\end{split}
\end{equation}

Based on (1), the Taylor expansion approximation of 
$D^{lk',n}_{k,n_H}$ can be expressed as 
$D^{lk',n}_{k,n_H} \approx  r^{Short}_{k,lk',n} - (n^x_Hd\cos \phi^{Short}_{k,lk',n}\cos \theta^{Short}_{k,lk',n}+n^z_Hd\sin \phi^{Short}_{k,lk',n})$.
Similarly, $D^{lk',n}_{k,n_R}$ can be approximated as $D^{lk',n}_{k,n_R} \approx  r^{Short}_{k,lk',n} - (n^x_Rd\cos \phi^{Short}_{k,lk',n}\cos \theta^{Short}_{k,lk',n}+n^z_Rd\sin \phi^{Short}_{k,lk',n})$.
Let us denote 
\begin{align}
\!\! \Psi^{Short}_{k,lk',n}\!=\!\cos \phi^{Short}_{k,lk',n}\!\cos \theta^{Short}_{k,lk',n}, \quad  \!\!\!\! 
\Omega^{Short}_{k,lk',n}\!=\! \sin \phi^{Short}_{k,lk',n}, \tag{24}
\end{align}
and then there is
\begin{equation}
	\begin{split}
		\begin{aligned}
			\label{27}
\!\!\!\!\!   \tau^{lk',n}_{k,n_{\!H}\!,n_{\!R}} \!\!=\!\frac{ 2r^{Short}_{k,lk'\!,n}\!\!-\!(n^x_{\!H}\!\!+\!\!n^x_{\!R})d\Psi^{Short}_{k,lk'\!,n}\!\!-\!(n^z_{\!H}\!\!+\!\!n^z_{\!R})\Omega^{Short}_{k,lk'\!,n}}{c}.
		\end{aligned}
	\end{split}\tag{25}
\end{equation}
Then (23) can be rewritten as
\begin{equation}
	\begin{split}
		\begin{aligned}
			\label{27}
&\!\!\!\!\!\! h^{lk',n,m}_{k,n_H,n_R} \!=\! \alpha^{lk'}_k e^{-j2\pi f_m \! \frac{2r^{Short}_{k,lk'\!,n}-(n^x_H+n^x_R)d\Psi^{Short}_{k,lk'\!,n}-(n^z_H+n^z_R)\Omega^{Short}_{k,lk'\!,n}}{c}}
\\&\!\!\!\!\!\! = \!\! \alpha^{lk'}_k\!\! e^{\!-\! j\! 4\pi\! f_m \!  \frac{r^{L\!o\!n\!g}_{k,l}}{c}}\!\!
e^{\!j\!4\pi\! f_m \!\frac{ v^{L\!o\!n\!g}_{r\!,k\!,l}\!n\!T_{s}}{c}}\!\!
e^{\!j2\pi\! f_m\!\!\frac{(\!n^x_H\!+ n^x_R\!)d\!\Psi^{Short}_{k\!,lk'\!,n} \!+\!(\! n^z_H+ n^z_R\!)d\Omega^{Short}_{k\!,lk'\!,n} }{c}}. 
		\end{aligned}
	\end{split}\tag{26}
\end{equation}
In narrowband OFDM systems, the Doppler squint effect and beam squint effect are typically negligible \cite{2023arXiv230512064L}, and thus (26) can be 
further represented as 
\begin{equation}
	\begin{split}
		\begin{aligned}
			\label{27}
h^{lk',n,m}_{k,n_H,n_R} & = \alpha^{lk'}_k e^{-j 4\pi f_m  \frac{r^{Long}_{k,l}}{c}}
e^{j4\pi f_0 \frac{ v^{Long}_{r,k,l}nT_{s}}{c}} \times
\\ & \kern 20pt 
e^{j2\pi f_0\frac{(n^x_H+ n^x_R)d\Psi^{Short}_{k,lk',n} +(n^z_H+ n^z_R)d\Omega^{Short}_{k,lk',n} }{c}}.
		\end{aligned}
	\end{split}\tag{27}
\end{equation}

We denote $\mathbf{H}^{lk'\!,n,m}_k \in \mathbb{C}^{N_R\times N_H}$
as the overall frequency-domain sensing echo channel matrix on the $m$-th subcarrier at the $n$-th OFDM symbol within the $(l,k')$-th UTS from the HU-UPA to the $k$-th dynamic  target and then back to the RU-UPA, whose $(n_R,n_H)$-th element is
$[\mathbf{H}^{lk'\!,n,m}_k]_{n_R,n_H} = h^{lk',n,m}_{k,n_H,n_R}$. 
Moreover, the matrix $\mathbf{H}^{lk'\!,n,m}_k$ can be decomposed as 
\begin{equation}
	\begin{split}
		\begin{aligned}
			\label{deqn_ex1a}
\!\!\!\! \mathbf{H}^{lk'\!,n,m}_k &\!=\! \alpha^{lk'}_k  e^{\!-j4\pi\! f_m \!  \frac{r^{Long}_{k,l}}{c}}\!
e^{j4\pi f_0\frac{v^{Long}_{r,k,l}nT_s}{c}} \times \\& \kern 16pt 
\mathbf{a}_{R}(\Psi^{Short}_{k,lk',n},\Omega^{Short}_{k,lk',n})
\mathbf{a}^T_{H}(\Psi^{Short}_{k,lk',n},\Omega^{Short}_{k,lk',n}),
		\end{aligned}
	\end{split}\tag{28}
\end{equation}
where $\mathbf{a}_{R}(\Psi,\Omega)$ is the array  steering vector for spatial-domain direction $(\Psi,\Omega)$ of RU-UPA with the form
\begin{equation}
	\begin{split}
		\begin{aligned}
			\label{deqn_ex1a}
			\mathbf{a}_{R}(\Psi,\Omega)\! =
			\mathbf{a}_{R}^x(\Psi)\otimes \mathbf{a}_{R}^z(\Omega)
			\in\! \mathbb{C}^{N_R\times 1}.
		\end{aligned}
	\end{split}\tag{29}
\end{equation}
Here $\otimes$ denotes the Kronecker product, and 
\begin{align}
	\!\!\mathbf{a}_{R}^x(\Psi)&\!=\![1,e^{j\frac{2\pi f_0d\Psi}{c}},...,e^{j\frac{2\pi f_0d\Psi}{c}(N_{R}^x-1)}]^T \!\!\in\! \mathbb{C}^{N_{R}^x\times 1} \tag{30},\\
	\!\!\mathbf{a}_{R}^z(\Omega)&\!=\! [1,e^{j\frac{2\pi f_0d\Omega}{c}},...,e^{j\frac{2\pi f_0d\Omega}{c}(N_{R}^z-1)}]^T \!\! \in\! \mathbb{C}^{N_{R}^z\times 1} \tag{31}.
\end{align}
Besides, $\alpha^{lk'}_k$
 is usually modeled as 
 $\alpha^{lk'}_k =\sqrt{\frac{\lambda^2}{(4\pi)^3 (r^{Long}_{k,l})^4}}\sigma_{k}$, and 
$\sigma_{k}$  is the radar cross section (RCS) of the $k$-th dynamic target.
Without loss of  generality, we  assume that the RCS follows the   Swerling $\rm \uppercase\expandafter{\romannumeral1}$ model. 

Based on (29), the  sensing echo channel of all $K$ dynamic targets on the $m$-th subcarrier of the $n$-th OFDM symbol in the $(l,k')$-th UTS can be represented as
\begin{equation}
	\begin{split}
		\begin{aligned}
			\label{deqn_ex1a}
\mathbf{H}^{target}_{lk',n,m} = \sum_{k=1}^K
\mathbf{H}^{lk',n,m}_k.
		\end{aligned}
	\end{split}\tag{32}
\end{equation}

In addition, since the real physical world is composed of dynamic targets and static environment,  
the RU-UPA  will receive both the effective echoes caused by interested dynamic targets (dynamic target echoes) and the undesired  echoes caused by uninterested background environment (clutter). 
Following our previous work \cite{2023arXiv231101674L}, we  model the static environmental clutter channel 
on the $m$-th subcarrier of the $n$-th OFDM symbol in the $(l,k')$-th UTS as
\begin{equation}
	\begin{split}
		\begin{aligned}
			\label{deqn_ex1a}
\!\!\!\!\! \mathbf{H}^{b\!a\!c\!k\!g\!r\!o\!u\!n\!d}_{lk',n,m} \!\!\!=\! \!\!\!
\sum_{i'=1}^{I'}\!\!\!\beta^{\mathfrak{c}\!,lk'}_{i'}
\!\!\!
e^{\!-\!j\!4\pi\! f_{\!m} \!\!  \frac{r^{\mathfrak{c}}_{i'\!\!,lk'}}{c}}\!\!
\mathbf{a}_{\!R}\!(\!\Psi^{\mathfrak{c}}_{i'\!\!,lk'}\!,\!\Omega^{\mathfrak{c}}_{i'\!\!,lk'}\!\!)
\mathbf{a}^T_{\!H}\!(\!\Psi^{\mathfrak{c}}_{i'\!\!,lk'}\!,\!\Omega^{\mathfrak{c}}_{i'\!\!,lk'}\!\!),
		\end{aligned}
	\end{split}\tag{33}
\end{equation}
where $I'$ is the total number of static environmental clutter scattering units,  $\Psi^{\mathfrak{c}}_{i'\!,lk'}=\cos\phi^{\mathfrak{c}}_{i'\!,lk'}\cos\theta^{\mathfrak{c}}_{i'\!,lk'}$, $\Omega^{\mathfrak{c}}_{i'\!,lk'}=\sin\phi^{\mathfrak{c}}_{i'\!,lk'}$, 
$(r^{\mathfrak{c}}_{i'\!,lk'},\theta^{\mathfrak{c}}_{i'\!,lk'},\phi^{\mathfrak{c}}_{i'\!,lk'})$ is the position of the $i'$-th  clutter scattering unit, $\beta^{\mathfrak{c},lk'}_{i'} = \sqrt{\frac{\lambda^2}{(4\pi)^3 (r^{\mathfrak{c}}_{i'\!,lk'})^4}}\sigma^{\mathfrak{c}}_{i'}$
is the channel fading factor, 
and $\sigma^{\mathfrak{c}}_{i'}$ is the RCS of the $i'$-th   clutter scattering unit that also 
follows the Swerling $\rm \uppercase\expandafter{\romannumeral1}$ model.
Due to the random distribution of clutter scattering units in various directions and distances, when the number of clutter scattering units $I'$ is large enough, $\mathbf{H}^{background}_{lk',n,m}$ can be  considered as a random channel. Therefore, a low complexity clutter channel generation method is to directly approximate $\mathbf{H}^{background}_{lk',n,m}$ as
\begin{equation}
	\begin{split}
		\begin{aligned}
			\label{deqn_ex1a}
\mathbf{H}^{background}_{lk',n,m} \approx 
\beta^{\mathfrak{c}}_{lk'} \mathbf{H}^{\mathfrak{c}}_{lk'\!,m}, 
		\end{aligned}
	\end{split}\tag{34}
\end{equation}
where $\mathbf{H}^{\mathfrak{c}}_{lk'\!,m}$  is a complex Gaussian matrix with the dimension of $N_R \times N_H$, and $\beta^{\mathfrak{c}}_{lk'}$  is the clutter power regulation factor.
Formulas (33) and (34) indicate that since the static  clutter scattering unit remains stationary for $N$ OFDM symbol times, $\mathbf{H}^{background}_{lk',n,m}$ would remain unchanged for $N$  symbols.

Based on (32) and (34), the overall sensing echo channel of  dynamic targets and static environment on the $m$-th subcarrier of the $n$-th OFDM symbol in the $(l,k')$-th UTS is 
\begin{equation}
	\begin{split}
		\begin{aligned}
			\label{deqn_ex1a}
\mathbf{H}^{sensing}_{lk',n,m} = 			\mathbf{H}^{target}_{lk',n,m} +\mathbf{H}^{background}_{lk',n,m}.
		\end{aligned}
	\end{split}\tag{35}
\end{equation}

\section{6D Radar Sensing and Tracking}

In this section,
we provide the sensing echo signals model,
and then  propose
a novel 6D sensing and tracking scheme for dynamic target sensing.

\vspace{-4mm}

\subsection{Echo Signals Model}

We design that the BS employs the Kalman filtering algorithm to track the dynamic targets within $L$  TTSs.
Assume that the predicted value of the 6D parameters for the $k'$-th dynamic target within the $l$-th TTS is $\widetilde{\mathbf{S}}^{Long}_{k',l}=[\widetilde{r}^{Long}_{k',l}, \widetilde{\theta}^{Long}_{k',l}, \widetilde{\phi}^{Long}_{k',l}, \widetilde{v}^{Long}_{r,k',l}, \widetilde{\omega}^{Long}_{\theta,k',l}, \widetilde{\omega}^{Long}_{\phi,k',l}]^T$.
Then, the HUUPA of the BS needs to generate one sensing beam towards $(\widetilde{\theta}^{Long}_{k',l}, \widetilde{\phi}^{Long}_{k',l})$ direction during the $(l,k')$-th UTS to re-sense the $k'$-th target.
Hence based on (6), the  transmission signals of HU-UPA during the $(l,k')$-th UTS can be expressed  as 
\begin{equation}
	\begin{split}
		\begin{aligned}
			\label{deqn_ex1a}
\mathbf{x}_{lk'\!\!,n,m} \!\!=\!\!  \sqrt{\frac{\!\grave{\rho}_{lk'}\!P_t}{N_H}}\mathbf{a}_{\!H}(\cos \widetilde{\phi}^{L\!o\!n\!g}_{k',l}\! \cos \widetilde{\theta}^{L\!o\!n\!g}_{k',l},\sin \widetilde{\phi}^{L\!o\!n\!g}_{k',l})s^{t,lk'}_{n,m}.
		\end{aligned}
	\end{split}\tag{36}
\end{equation}
Then the sensing echo signals on the $m$-th subcarrier of the $n$-th OFDM symbol received by the RU-UPA in the $(l,k')$-th UTS can be represented as
\begin{equation}
	\begin{split}
		\begin{aligned}
			\label{deqn_ex1a}
&\mathbf{y}^{lk'}_{n,m} \!=\!  \mathbf{H}^{sensing}_{lk',n,m}  \mathbf{x}_{lk'\!,n,m}^* \!+\! \mathbf{n}^{lk'}_{n,m} \\& =  \mathbf{H}^{target}_{lk',n,m} \mathbf{x}_{lk'\!,n,m}^* \!+\! \mathbf{H}^{background}_{lk',n,m} \mathbf{x}_{lk'\!,n,m}^* \!+\! \mathbf{n}^{lk'}_{n,m}, 
\\&\kern 55pt  n=0,...,N-1,\kern 4pt m=0,...,M-1,
		\end{aligned}
	\end{split}\tag{37}
\end{equation}
where $[\mathbf{n}^{lk'}_{n,m}]_{n_R}$ is the zero-mean additive   Gaussian   noise  with variance  $\sigma_{lk'}^2$.
Note that $\mathbf{y}^{lk'}_{n,m}$ represents the echo signals received by all $N_R=N_R^x\times N_R^z$ receiving antennas, and then we can reformat the  vector $\mathbf{y}^{lk'}_{n,m}$ into matrix form as $\mathbf{Y}^{lk'}_{n,m}$
\begin{equation}
	\begin{split}
		\begin{aligned}
			\label{deqn_ex1a}
&\mathbf{Y}^{lk'}_{n,m} =  {\rm reshape} \{\mathbf{y}^{lk'}_{n,m},[N_R^z,N_R^x]\}
\in \mathbb{C}^{N_{R}^z\times N_{R}^x}.
		\end{aligned}
	\end{split}\tag{38}
\end{equation}
Furthermore, we can stack $\mathbf{Y}^{lk'}_{n,m}$ into one echoes tensor 
${\mathbf{Y}}_{cube}^{lk'} \in \mathbb{C}^{N_{R}^z\times N_{R}^x \times N\times M}$, 
whose $(n^z_R,n^x_R,n,m)$-th element is ${\mathbf{Y}}_{cube}^{lk'}[n^z_R,n^x_R,n,m] = [\mathbf{Y}^{lk'}_{n,m}]_{n^z_R,n^x_R}$.

Note that ${\mathbf{Y}}_{cube}^{lk'}$ includes the sensing  channel $\mathbf{H}^{sensing}_{lk',n,m}$, transmitting beamforming, and transmission symbols $s^{t,lk'}_{n,m}$, while   targets sensing can be understood as an estimation of $\mathbf{H}^{sensing}_{lk',n,m}$. However, random transmission symbols would affect the estimation of  sensing  channel, and thus we need to  erase the transmission symbols from the received signals to obtain  equivalent echo  channel (EEC).
Specifically, the EEC corresponding to $\mathbf{Y}^{lk'}_{n,m}$ can be obtained as
$\tilde{\mathbf{Y}}^{lk'}_{n,m} = \mathbf{Y}^{lk'}_{n,m}/s^{t,lk'}_{n,m}$.
Then we can stack $\tilde{\mathbf{Y}}^{lk'}_{n,m}$ 
into an EEC tensor $\tilde{\mathbf{Y}}_{cube}^{lk'} \in \mathbb{C}^{N_{R}^z\times N_{R}^x \times N\times M}$
with $\tilde{\mathbf{Y}}_{cube}^{lk'}[n^z_R,n^x_R,n,m] = [\tilde{\mathbf{Y}}^{lk'}_{n,m}]_{n^z_R,n^x_R}$. 

\vspace{-2mm}

\subsection{Static Environmental Clutter Filtering}

It can be analyzed from (35) and (37) that
the echo signals ${\mathbf{Y}}_{cube}^{lk'}$ includes both dynamic target echoes and static environment echoes, 
and the  EEC $\tilde{\mathbf{Y}}_{cube}^{lk'}$  also includes both the
EEC of  dynamic targets (DT-EEC) and the
EEC of  static environment (SE-EEC).
When we focus on dynamic target sensing, the SE-EEC in  original echo signals would cause negative interference to dynamic target sensing, and thus  SE-EEC can be referred to as  clutter-EEC.
To address this negative interference,  we need  to filter out the interference of
clutter-EEC  and  to extract the  effective DT-EEC from $\tilde{\mathbf{Y}}_{cube}^{lk'}$.

While the environmental clutter filtering is  necessary in sensing processing,  it is not the focus of this work. According to the clutter suppression method  in \cite{2023arXiv231101674L}, we may express the effective DT-EEC after static clutter filtering as $\check{\mathbf{Y}}_{cube}^{lk'}$,
whose $[:,:,n,m]$-th sub-matrix is
$\check{\mathbf{Y}}_{cube}^{lk'}[:,:,n,m]=
\check{\mathbf{Y}}^{lk'}_{n,m} = {\rm reshape} \{\check{\mathbf{y}}^{lk'}_{n,m},[N_R^z,N_R^x]\}$
with  
\begin{equation}
	\begin{split}
		\begin{aligned}
			\label{deqn_ex1a}
 \check{\mathbf{y}}^{lk'}_{n,m}  & \approx   \mathbf{H}^{target}_{lk',n,m} \mathbf{x}_{lk'\!,n,m}^*/s^{t,lk'}_{n,m} \!+\!  \check{\mathbf{n}}^{lk'}_{n,m}
\\& = \sum_{k=1}^K \mathbf{H}^{lk',n,m}_k \mathbf{x}_{lk'\!,n,m}^*/s^{t,lk'}_{n,m} \!+\!  \check{\mathbf{n}}^{lk'}_{n,m}, 
		\end{aligned}
	\end{split}\tag{39}
\end{equation}
where $\check{\mathbf{n}}^{lk'}_{n,m}$  is the noise after static clutter filtering.

\subsection{Echo Signals Analysis}

\begin{figure*}[!t]
	\normalsize
	\setcounter{MYtempeqncnt}{\value{equation}}
	\vspace*{4pt}
	\setcounter{equation}{40}
	\begin{equation}
		\begin{split}
			\begin{aligned}
				\label{deqn_ex1a}
				\check{\mathbf{y}}^{lk'}_{n\!,m} \!\!=\!\!\!\!  \sum_{k=1}^K \! \! 
				\left[ \! \alpha^{lk'}_k \!\! e^{\!-\!j4\pi\! f_m \!  \frac{r^{L\!o\!n\!g}_{k,l}}{c}}\!
				e^{j4\pi \!f_0\!\frac{v^{L\!o\!n\!g}_{\!r,k,l}n\!T_s}{c}}\!
				\mathbf{a}_{\!R}(\!\Psi^{Short}_{k,lk'\!,n},\Omega^{Short}_{k,lk'\!,n}\!)
				\mathbf{a}^T_{\!H}(\!\Psi^{Short}_{k,lk'\!,n},\Omega^{Short}_{k,lk'\!,n}\!)\!  \sqrt{\!\frac{\grave{\rho}_{lk'}\!P_t}{N_H}}\!
				\mathbf{a}^*_{\!H}\!(\!\cos \! \widetilde{\phi}^{L\!o\!n\!g}_{k',l} \cos \widetilde{\theta}^{L\!o\!n\!g}_{k',l},\sin\! \widetilde{\phi}^{L\!o\!n\!g}_{k',l})\!\right]
				\!\!\!+\!  \check{\mathbf{n}}^{lk'}_{n\!,m}.
			\end{aligned}
		\end{split}\tag{40}
	\end{equation}
	\begin{equation}
		\begin{split}
			\begin{aligned}
				\label{deqn_ex1a}
				\check{\mathbf{y}}^{lk'}_{n\!,m} &\!\!=\!\!\!\!  \sum_{k=1}^K \! \! 
				\left[ \! \alpha^{lk'}_k \!\! e^{\!-\!j4\pi\! f_m \!  \frac{r^{L\!o\!n\!g}_{k,l}}{c}}\!
				e^{j4\pi \!f_0\!\frac{v^{L\!o\!n\!g}_{\!r,k,l}n\!T_s}{c}}\!
				\mathbf{a}_{\!R}(\!\Psi^{Short}_{k,lk'\!,n},\Omega^{Short}_{k,lk'\!,n}\!)
				\mathbf{a}^T_{\!H}(\!\Psi^{Short}_{k,lk'\!,n},\Omega^{Short}_{k,lk'\!,n}\!)\!  \sqrt{\!\frac{\grave{\rho}_{lk'}\!P_t}{N_H}}\!
				\mathbf{a}^*_{\!H}\!(\!\Psi^{Short}_{k'\!,lk'\!,0},\Omega^{Short}_{k'\!,lk'\!,0}\!)\!\right]
				\!\!\!+\!  \check{\mathbf{n}}^{lk'}_{n\!,m}
				\\& \!\! = \alpha^{lk'}_{k'} \!\! e^{\!-\!j4\pi\! f_m \!  \frac{r^{L\!o\!n\!g}_{k',l}}{c}}\!
				e^{j4\pi \!f_0\!\frac{v^{L\!o\!n\!g}_{\!r,k',l}n\!T_s}{c}}\!
				\mathbf{a}_{\!R}(\!\Psi^{Short}_{k',lk'\!,n},\Omega^{Short}_{k',lk'\!,n}\!)
				\mathbf{a}^T_{\!H}(\!\Psi^{Short}_{k',lk'\!,n},\Omega^{Short}_{k',lk'\!,n}\!)\!  \sqrt{\!\frac{\grave{\rho}_{lk'}\!P_t}{N_H}}\!
				\mathbf{a}^*_{\!H}\!(\!\Psi^{Short}_{k'\!,lk'\!,0},\Omega^{Short}_{k'\!,lk'\!,0}\!) \!+\!  \check{\mathbf{n}}^{lk'}_{n\!,m}
				\\&\!\! = \mathscr{G}^{lk'}_{k'}
				e^{\!-\!j4\pi\! f_m \!  \frac{r^{L\!o\!n\!g}_{k',l}}{c}}\!
				e^{j4\pi \!f_0\!\frac{v^{L\!o\!n\!g}_{\!r,k',l}n\!T_s}{c}}
				e^{j\!\frac{\pi\! f_0d(\Omega^{Short}_{k'\!,lk'\!,n}\!\!-\Omega^{Short}_{k'\!,lk'\!,0})(N_H^z\!-\!1)}{c}}
				e^{j\!\frac{\pi\! f_0d(\Psi^{Short}_{k'\!,lk'\!,n}\!\!-\Psi^{Short}_{k'\!,lk'\!,0})(N_H^x\!-\!1)}{c}}
				\mathbf{a}_{\!R}(\!\Psi^{Short}_{k'\!,lk'\!,n},\Omega^{Short}_{k'\!,lk'\!,n}\!) \!+\!  \check{\mathbf{n}}^{lk'}_{n\!,m}.
			\end{aligned}
		\end{split}\tag{41}
	\end{equation}
	\begin{equation}
		\begin{split}
			\begin{aligned}
				\label{deqn_ex1a}
& \check{y}^{lk'}_{n,m,n^z_R,n^x_R} = \check{\mathbf{Y}}_{cube}^{lk'}[n^z_R,n^x_R,n,m] \!\!= \\ &\!  \mathscr{G}^{lk'}_{k'} \!\! 
e^{\!-\!j4\pi\! f_m \!  \frac{r^{L\!o\!n\!g}_{k',l}}{c}}\!
e^{j4\pi \!f_0\!\frac{v^{L\!o\!n\!g}_{\!r,k',l}\! n\!T_s}{c}}\!\!
e^{j\!\frac{\pi\! f_0d(\Omega^{Short}_{k'\!,lk'\!,n}\!\!-\Omega^{Short}_{k'\!,lk'\!,0}\!)(N_H^z\!-\!1)}{c}}\!\!
e^{j\!\frac{\pi\! f_0d(\Psi^{Short}_{k'\!,lk'\!,n}\!\!-\Psi^{Short}_{k'\!,lk'\!,0})(N_H^x\!-\!1)\!}{c}}\!\!
e^{j\!\frac{2\pi\! f_0n_R^zd\Omega^{Short}_{k'\!,lk'\!,n}}{c}}\!\! 
e^{j\!\frac{2\pi\! f_0n_R^xd\Psi^{Short}_{k'\!,lk'\!,n}}{c}}
\!\!+\!  \check{n}^{lk'}_{n\!,m\!,n^z_R\!,n^x_R}.
			\end{aligned}
		\end{split}\tag{42}
	\end{equation}
\begin{equation}
	\begin{split}
		\begin{aligned}
			\label{deqn_ex1a}
&\Psi^{Short}_{k'\!,lk'\!,n}\!\!-\Psi^{Short}_{k'\!,lk'\!,0} = \cos\phi^{Short}_{k'\!,lk'\!,n}\cos\theta^{Short}_{k'\!,lk'\!,n} -  \cos\phi^{Short}_{k'\!,lk'\!,0}\cos\theta^{Short}_{k'\!,lk'\!,0} \\& = -[\cos\phi^{Long}_{k',l}\cos\theta^{Long}_{k',l} - \cos(\phi^{Long}_{k',l} - \omega^{Long}_{\phi,k',l}nT_{s})\cos(\theta^{Long}_{k',l}-\omega^{Long}_{\theta,k',l}nT_{s})] \\& = -
\frac{\cos\phi^{Long}_{k',l}\cos\theta^{Long}_{k',l}- \cos(\phi^{Long}_{k',l} - \frac{\omega^{Long}_{\phi,k',l}}{\mathscr{W}_{k'\!,l}}nT_{s}\mathscr{W}_{k'\!,l})\cos(\theta^{Long}_{k',l}-
\frac{\omega^{Long}_{\theta,k',l}}{\mathscr{W}_{k'\!,l}}nT_{s}\mathscr{W}_{k'\!,l})}{nT_s\mathscr{W}_{k'\!,l}}nT_s\mathscr{W}_{k'\!,l} \\&\approx -
\left[-\sin\phi^{Long}_{k',l}\cos\theta^{Long}_{k',l}\frac{\omega^{Long}_{\phi,k',l}}{\mathscr{W}_{k'\!,l}} - \cos\phi^{Long}_{k',l}\sin\theta^{Long}_{k',l}
\frac{\omega^{Long}_{\theta,k',l}}{\mathscr{W}_{k'\!,l}}\right]nT_s\mathscr{W}_{k'\!,l}\\&=
\sin\phi^{Long}_{k',l}\cos\theta^{Long}_{k',l}\omega^{Long}_{\phi,k',l}nT_s+ \cos\phi^{Long}_{k',l}\sin\theta^{Long}_{k',l}
\omega^{Long}_{\theta,k',l}nT_s.
		\end{aligned}
	\end{split}\tag{44}
\end{equation}
	\setcounter{equation}{\value{MYtempeqncnt}}
\end{figure*}

Based on (28), (32) and (36),  
$\check{\mathbf{y}}^{lk'}_{n,m}$ in (39) can be calculated as (40) at the top of  next page. 
When the Kalman filter predicts accurately,
based on (16) and (17),
there should be 
$(\cos \! \widetilde{\phi}^{Long}_{k',l} \cos \widetilde{\theta}^{Long}_{k',l},\sin\! \widetilde{\phi}^{Long}_{k',l})= 
(\cos \! \phi^{Short}_{k'\!,lk'\!,0} \cos \theta^{Short}_{k'\!,lk'\!,0},\sin\! \phi^{Short}_{k'\!,lk'\!,0})
=(\Psi^{Short}_{k'\!,lk'\!,0},\Omega^{Short}_{k'\!,lk'\!,0})$.
Then in massive MIMO system, due to $K$ dynamic targets owning different directions, (40) can be calculated as (41) at the top of  next page,
where $\mathscr{G}^{lk'}_{k'} = \alpha^{lk'}_{k'}\!\!\!  \sqrt{\!\!\frac{\grave{\rho}_{lk\!'}\!\! P_t}{N_H}}
\!\frac{\sin [\!\frac{\pi\!f_0d}{c} (\Omega^{Short}_{k'\!,lk'\!,n}\!\!-\Omega^{Short}_{k'\!,lk'\!,0})\!N_H^z\!]}
{\sin [\frac{\pi\!f_0d}{c} (\Omega^{Short}_{k'\!,lk'\!,n}\!\!-\Omega^{Short}_{k'\!,lk'\!,0})]}
\!\frac{\sin [\!\frac{\pi\!f_0d}{c} (\Psi^{Short}_{k'\!,lk'\!,n}\!\!-\Psi^{Short}_{k'\!,lk'\!,0})\!N_H^x\!]}
{\sin [\frac{\pi\!f_0d}{c} (\Psi^{Short}_{k'\!,lk'\!,n}\!\!-\Psi^{Short}_{k'\!,lk'\!,0})]}$.
Next, based on   (41),
the $[n^z_R,n^x_R,n,m]$-th element in 
$\check{\mathbf{Y}}_{cube}^{lk'}$ can be 
expressed as (42) at the top of next page.

To further simplify (42), we note that
$\phi^{Short}_{k,lk',0} =\phi^{Long}_{k,l}$ (based on (17)) and find that 
\begin{equation}
	\begin{split}
		\begin{aligned}
			\label{deqn_ex1a}
&\Omega^{Short}_{k'\!,lk'\!,n}\!\!-\Omega^{Short}_{k'\!,lk'\!,0} = \sin \phi^{Short}_{k'\!,lk'\!,n} -
\sin \phi^{Short}_{k'\!,lk'\!,0} \\& = 
- \frac{\sin \phi^{Short}_{k'\!,lk'\!,0} -
\sin (\phi^{Short}_{k'\!,lk'\!,0}-\omega^{Long}_{\phi,k',l}nT_{s} )}{\omega^{Long}_{\phi,k',l}nT_{s}}\omega^{Long}_{\phi,k',l}nT_{s} \\& =
- \frac{\sin \phi^{Long}_{k',l} -
\sin (\phi^{Long}_{k',l}-\omega^{Long}_{\phi,k',l}nT_{s} )}{\omega^{Long}_{\phi,k',l}nT_{s}}\omega^{Long}_{\phi,k',l}nT_{s} \\& \approx
- \cos \phi^{Long}_{k',l}\omega^{Long}_{\phi,k',l}nT_{s}.
		\end{aligned}
	\end{split}\tag{43}
\end{equation}
Besides, we also note that $\theta^{Short}_{k,lk',0}=\theta^{Long}_{k,l}$ (based on (16)).
According to the definition and properties of directional derivatives of binary function, we 
compute $\Psi^{Short}_{k'\!,lk'\!,n}\!\!-\Psi^{Short}_{k'\!,lk'\!,0}$ as shown in (44) at the top of this page, where $\mathscr{W}_{k'\!,l} = \sqrt{(\omega^{Long}_{\phi,k',l})^2+(\omega^{Long}_{\theta,k',l})^2}$.

Based on (43) and (44),
the $[n^z_R,n^x_R,n,m]$-th element in 
$\check{\mathbf{Y}}_{cube}^{lk'}$, i.e., 
the $\check{y}^{lk'}_{n,m,n^z_R,n^x_R}$  in (42) 
can be  calculated as shown in (45) at the top of this page.
Formula (45) indicates that $\check{y}^{lk'}_{n,m,n^z_R,n^x_R}$ includes the distance term, pitch angle term, horizontal angle term, radial velocity term, pitch angular velocity term, and horizontal angular velocity term. Therefore, it is definite	 to estimate the 6D motion parameters of dynamic targets from DT-EEC $\check{\mathbf{Y}}_{cube}^{lk'}$ and realize 6D radar sensing.

\newcounter{TempEqCnt} 
\setcounter{TempEqCnt}{\value{equation}} 
\setcounter{equation}{4} 
\begin{figure*}[ht] 
	\begin{equation}
		\begin{split}
			\begin{aligned}
				\label{deqn_ex1a}
\check{y}^{lk'}_{n,m,n^z_R,n^x_R} = &
\mathscr{G}^{lk'}_{k'}\!\! 
\underbrace{e^{\!-\!j4\pi\! f_m \!\!  \frac{r^{L\!o\!n\!g}_{k',l}}{c}}\!\!}_{\rm distance} \underbrace{
e^{j\!\frac{2\pi\! f_0d}{c}\!\sin\!\phi^{Long}_{k'\!,l}n^z_R}\!}_{\rm pitch \kern 2pt angle} \underbrace{
e^{j\!\frac{2\pi\! f_0d}{c}\!\cos\phi^{Long}_{k',l}\!\cos\!\theta^{Long}_{k',l}n^x_R}}_{\rm horizontal \kern 2pt angle}
\times  \underbrace{e^{j4\pi \!f_0\!\frac{v^{L\!o\!n\!g}_{\!r,k',l}n\!T_s}{c}}}_{\rm radial \kern 2pt velocity} \times \\&
\underbrace{e^{\!-j\!\frac{\pi\! f_0d}{c}[(N_H^z\!-\!1)\!\cos\!\phi^{L\!o\!n\!g}_{k',l}-(\!N^x_H\!-\!1)\!\sin\!\phi^{L\!o\!n\!g}_{k',l}\!\cos\!\theta^{L\!o\!n\!g}_{k'\!,l}]\omega_{\phi,k'\!,l}^{L\!o\!n\!g}nT_s}
e^{\!-j\!\frac{2\pi f_0d}{c}\!(n_R^z\cos\phi^{Long}_{k',l}-n^x_R\sin\phi^{Long}_{k',l}\cos\theta^{Long}_{k',l})\omega_{\phi,k'\!,l}^{L\!o\!n\!g}nT_s}}_{\rm pitch \kern 2pt angular \kern 2pt velocity} \times \\&
\underbrace{e^{j\frac{\pi f_0d(N_H^x-1)}{c}\cos\phi^{Long}_{k',l}\sin\theta^{Long}_{k',l}\omega^{Long}_{\theta,k',l}nT_s}
e^{j\frac{2\pi f_0 n_R^xd}{c}\cos\phi^{Long}_{k',l}\sin\theta^{Long}_{k'.l}\omega^{Long}_{\theta,k',l}nT_s}}_{\rm horizontal \kern 2pt angular \kern 2pt velocity}
+ \check{n}^{lk'}_{n,m,n^z_R,n^x_R}
			\end{aligned}
		\end{split}\tag{45}
	\end{equation}
	\hrulefill 
	\vspace*{8pt} 
\end{figure*}

\vspace{-4.4mm}

\subsection{Angle Direction and Distance Estimation}

Let us transform $\check{\mathbf{Y}}_{cube}^{lk'} \in \mathbb{C}^{N_{R}^z\times N_{R}^x \times N\times M}$ into an $\Omega$-matrix $\mathbf{Y}^{lk'}_{\Omega}$ with the dimension of $N_R^z\times N_R^xNM$. Based on (45), $\mathbf{Y}^{lk'}_{\Omega}$  can be represented as
\begin{equation}
	\begin{split}
		\begin{aligned}
			\label{deqn_ex1a}
\mathbf{Y}^{lk'}_{\Omega}= \mathbf{k}^{lk'}_{\Omega}(\Omega^{Long}_{k',l})\cdot
\mathbf{x}^{lk'}_{\Omega} + \mathbf{N}^{lk'}_{\Omega} \in \mathbb{C}^{N_R^z\times N_{R}^x  N M},
		\end{aligned}
	\end{split}\tag{46}
\end{equation}
where $\Omega^{Long}_{k',l} = \sin\phi^{Long}_{k',l}$, $\mathbf{k}^{lk'}_{\Omega}(\Omega) = [1,e^{j\frac{2\pi f_0d\Omega}{c}},...,e^{j\frac{2\pi f_0d\Omega}{c}(N_{R}^z-1)}]^T \!\! \in\! \mathbb{C}^{N_{R}^z\times 1}$ is defined as the \emph{second spatial-domain direction array steering vector}, $\mathbf{x}^{lk'}_{\Omega}\in \mathbb{C}^{1\times N_{R}^x  N M}$ and $\mathbf{N}^{lk'}_{\Omega} \in \mathbb{C}^{N_R^z\times N_{R}^x  N M}$.
Since $\mathbf{Y}^{lk'}_{\Omega}$ is the array signals form related to the second spatial-domain direction array, we can estimate $\Omega^{Long}_{k',l}$ from $\mathbf{Y}^{lk'}_{\Omega}$ by utilizing array signal processing methods.

Here we adopt the
estimating signal
parameters via rotational variation techniques
(ESPRIT) method for parameter estimation\cite{32276,1164935,671426}.
Specifically, the covariance matrix of $\mathbf{Y}^{lk'}_{\Omega}$ can be calculated as
$\mathbf{R}_{\Omega}^{lk'} = \frac{1}{N^x_RNM}\mathbf{Y}^{lk'}_{\Omega}(\mathbf{Y}^{lk'}_{\Omega})^H$. 
We perform eigenvalue decomposition of  $\mathbf{R}_{\Omega}^{lk'}$  to obtain the diagonal matrix with eigenvalues ranging from large to small ($\mathbf{\Sigma}_{\Omega}^{lk'}$) and the corresponding eigenvector matrix ($\mathbf{U}_{\Omega}^{lk'}$), that is, $[\mathbf{U}_{\Omega}^{lk'}, \mathbf{\Sigma}_{\Omega}^{lk'}]={\rm eig}(\mathbf{R}_{\Omega}^{lk'})$.
Then  the minimum description length (MDL) criterion is utilized to estimate the number of dynamic targets from $\mathbf{\Sigma}_{\Omega}^{lk'}$  as $K_{lk'}^{\Omega}$\cite{mdl,mdl2}. We  extract the  parallel signal subspaces from $\mathbf{U}_{\Omega}^{lk'}$ as 
$\mathbf{U}_{\Omega,1}^{lk'} = \left[ \mathbf{U}_{\Omega}^{lk'}[1:N_R^z-1,1:K_{lk'}^{\Omega}], \mathbf{U}_{\Omega}^{lk'}[2:N_R^z,1:K_{lk'}^{\Omega}] \right] \in \mathbb{C}^{(N_R^z-1)\times 2K_{lk'}^{\Omega}}$ and compute 
$\tilde{\mathbf{R}}_{\Omega}^{lk'} = (\mathbf{U}_{\Omega,1}^{lk'})^H\mathbf{U}_{\Omega,1}^{lk'}\in\mathbb{C}^{2K_{lk'}^{\Omega}\times 2K_{lk'}^{\Omega}}$. Then we perform eigenvalue decomposition of 
$\tilde{\mathbf{R}}_{\Omega}^{lk'}$ to obtain the diagonal matrix with eigenvalues ranging from large to small ($\tilde{\mathbf{\Sigma}}_{\Omega}^{lk'}$) and the corresponding eigenvector matrix ($\tilde{\mathbf{U}}_{\Omega}^{lk'}$), that is, $[\tilde{\mathbf{U}}_{\Omega}^{lk'}, \tilde{\mathbf{\Sigma}}_{\Omega}^{lk'}]={\rm eig}(\tilde{\mathbf{R}}_{\Omega}^{lk'})$.
We extract $\tilde{\mathbf{U}}_{\Omega,a}^{lk'} = \tilde{\mathbf{U}}_{\Omega}^{lk'}[1:K_{lk'}^{\Omega},K_{lk'}^{\Omega}+1:2K_{lk'}^{\Omega}] \in \mathbb{C}^{K_{lk'}^{\Omega}\times K_{lk'}^{\Omega}}$ and 
$\tilde{\mathbf{U}}_{\Omega,b}^{lk'} = \tilde{\mathbf{U}}_{\Omega}^{lk'}[K_{lk'}^{\Omega}+1:2K_{lk'}^{\Omega},K_{lk'}^{\Omega}+1:2K_{lk'}^{\Omega}] \in \mathbb{C}^{K_{lk'}^{\Omega}\times K_{lk'}^{\Omega}}$,
and we compute $\check{\mathbf{R}}_{\Omega}^{lk'}=-\tilde{\mathbf{U}}_{\Omega,a}^{lk'}(\tilde{\mathbf{U}}_{\Omega,b}^{lk'})^{-1}$. Next, we perform eigenvalue decomposition of  $\check{\mathbf{R}}_{\Omega}^{lk'}$  to obtain the diagonal matrix with eigenvalues ranging from large to small ($\check{\mathbf{\Sigma}}_{\Omega}^{lk'}$) and the corresponding eigenvector matrix ($\check{\mathbf{U}}_{\Omega}^{lk'}$), that is, $[\check{\mathbf{U}}_{\Omega}^{lk'}, \check{\mathbf{\Sigma}}_{\Omega}^{lk'}]={\rm eig}(\check{\mathbf{R}}_{\Omega}^{lk'})$.
We take out the elements on the main diagonal of $\check{\mathbf{\Sigma}}_{\Omega}^{lk'}$ to form one  eigenvalues set as $\{\lambda_{\Omega,1}^{lk'},\lambda_{\Omega,2}^{lk'},...,\lambda_{\Omega,K_{lk'}^{\Omega}}^{lk'}\}$,
and  compute the \emph{space values} as
$\kappa_{\Omega,i}^{lk'} =\arctan{\frac{{\rm Imag}(\lambda_{\Omega,i}^{lk'})}{{\rm Real}(\lambda_{\Omega,i}^{lk'})}}$, where 
$i=1,2,...,K_{lk'}^{\Omega}$.
Since $\check{\mathbf{Y}}_{cube}^{lk'}$ only contains one dynamic target, there should be 
$K_{lk'}^{\Omega}=1$, and thus we  abbreviate the space value corresponding to $\mathbf{Y}^{lk'}_{\Omega}$ as $\kappa_{\Omega}^{lk'}$.
 Then the second spatial-domain direction of the $k'$-th dynamic target within the $l$-th TTS can be estimated as 
\begin{equation}
	\begin{split}
		\begin{aligned}
			\label{deqn_ex1a}
\check{\Omega}^{Long}_{k',l} =\frac{c\kappa_{\Omega}^{lk'}}{2\pi f_0 d}.
		\end{aligned}
	\end{split}\tag{47}
\end{equation}
Finally, the pitch angle of the $k'$-th dynamic target within the $l$-th TTS can be estimated as 
\begin{equation}
	\begin{split}
		\begin{aligned}
			\label{deqn_ex1a}
\check{\phi}_{k',l}^{Long} = \arcsin \left(		\check{\Omega}^{Long}_{k',l}\right)	 =\arcsin \left(	\frac{c\kappa_{\Omega}^{lk'}}{2\pi f_0 d}\right).
		\end{aligned}
	\end{split}\tag{48}
\end{equation}

\setcounter{TempEqCnt}{\value{equation}} 
\setcounter{equation}{4} 
\begin{figure*}[ht] 
	\begin{equation}
		\begin{split}
			\begin{aligned}
				\label{deqn_ex1a}
				&v^{Long,vir}_{k',l,n^x_R,n^z_R} = v^{Long}_{r,k',l}
				-\frac{d}{4}\left[(N_H^z-1)\cos\phi^{Long}_{k',l}
				-(N_H^x-1)\sin\phi^{Long}_{k',l}\cos\theta^{Long}_{k',l}\right]\omega^{Long}_{\phi,k',l} \\&-\!
				\frac{d}{2}(n_R^z\cos\phi^{Long}_{k',l}\!-\!n^x_R\sin\phi^{Long}_{k',l}\cos\theta^{Long}_{k',l})\omega^{Long}_{\phi,k',l}\!+\!
				\frac{d}{4}
				(N^x_H\!-\!1)\cos\phi^{Long}_{k',l}\sin\theta^{Long}_{k',l}\omega^{Long}_{\theta,k',l}
				\!+\!\frac{d}{2}n^x_R\cos\phi^{Long}_{k',l}\sin\theta^{Long}_{k',l}\omega^{Long}_{\theta,k',l}.
			\end{aligned}
		\end{split}\tag{54}
	\end{equation}
	\begin{equation}
		\begin{split}
			\begin{aligned}
				\label{deqn_ex1a}
				\check{y}^{lk'}_{n,m,n^z_R,n^x_R} = &
				\mathscr{G}^{lk'}_{k'}\!\! 
				\underbrace{e^{\!-\!j4\pi\! f_m \!\!  \frac{r^{L\!o\!n\!g}_{k',l}}{c}}\!\!}_{\rm distance} \underbrace{
					e^{j\!\frac{2\pi\! f_0d}{c}\!\sin\!\phi^{Long}_{k'\!,l}n^z_R}\!}_{\rm pitch \kern 2pt angle} \underbrace{
					e^{j\!\frac{2\pi\! f_0d}{c}\!\cos\phi^{Long}_{k',l}\!\cos\!\theta^{Long}_{k',l}n^x_R}}_{\rm horizontal \kern 2pt angle} \underbrace{e^{j4\pi \!f_0\!\frac{v^{Long,vir}_{k',l,n^x_R,n^z_R}n\!T_s}{c}}}_{\rm virtual-velocity} 
				+ \check{n}^{lk'}_{n,m,n^z_R,n^x_R}
			\end{aligned}
		\end{split}\tag{55}
	\end{equation}
	\hrulefill 
	\vspace*{8pt} 
\end{figure*}

Similarly, let us transform $\check{\mathbf{Y}}_{cube}^{lk'} \in \mathbb{C}^{N_{R}^z\times N_{R}^x \times N\times M}$ into an $\Psi$-matrix $\mathbf{Y}^{lk'}_{\Psi}$ with the dimension of $N_R^x\times N_R^zNM$. Based on (45), $\mathbf{Y}^{lk'}_{\Psi}$  can be represented as
\begin{equation}
	\begin{split}
		\begin{aligned}
			\label{deqn_ex1a}
			\mathbf{Y}^{lk'}_{\Psi}= \mathbf{k}^{lk'}_{\Psi}(\Psi^{Long}_{k',l})\cdot
			\mathbf{x}^{lk'}_{\Psi} + \mathbf{N}^{lk'}_{\Psi} \in \mathbb{C}^{N_R^x\times N_{R}^z  N M},
		\end{aligned}
	\end{split}\tag{49}
\end{equation}
where $\Psi^{Long}_{k',l} = \cos\phi^{Long}_{k',l}\cos\theta^{Long}_{k',l}$, $\mathbf{k}^{lk'}_{\Psi}(\Psi) = [1,e^{j\frac{2\pi f_0d\Psi}{c}},...,e^{j\frac{2\pi f_0d\Psi}{c}(N_{R}^x-1)}]^T \!\! \in\! \mathbb{C}^{N_{R}^x\times 1}$ is defined as the \emph{first spatial-domain direction array steering vector}, $\mathbf{x}^{lk'}_{\Psi}\in \mathbb{C}^{1\times N_{R}^z  N M}$ and $\mathbf{N}^{lk'}_{\Psi} \in \mathbb{C}^{N_R^x\times N_{R}^z  N M}$.
Since $\mathbf{Y}^{lk'}_{\Psi}$ is the array signals form related to the first spatial-domain direction array, we can estimate $\Psi^{Long}_{k',l}$ from $\mathbf{Y}^{lk'}_{\Psi}$ by utilizing array signal processing methods.
Similarly, we can employ the ESPRIT method to obtain the space value  corresponding to $\mathbf{Y}^{lk'}_{\Psi}$ as $\kappa_{\Psi}^{lk'}$.
Then the first spatial-domain direction of the $k'$-th dynamic target within the $l$-th TTS can be estimated as 
\begin{equation}
	\begin{split}
		\begin{aligned}
			\label{deqn_ex1a}
			\check{\Psi}^{Long}_{k',l} =\frac{c\kappa_{\Psi}^{lk'}}{2\pi f_0 d}.
		\end{aligned}
	\end{split}\tag{50}
\end{equation}
Then the horizontal angle  of the $k'$-th dynamic target within the $l$-th TTS can be estimated as 
\begin{equation}
	\begin{split}
		\begin{aligned}
			\label{deqn_ex1a}
\check{\theta}_{k',l}^{Long} = \arccos \left(		\frac{\check{\Psi}^{Long}_{k',l}}{\cos \check{\phi}_{k',l}^{Long}}  \right).
		\end{aligned}
	\end{split}\tag{51}
\end{equation}

To estimate the distance of the target,
let us   transform $\check{\mathbf{Y}}_{cube}^{lk'} \in \mathbb{C}^{N_{R}^z\times N_{R}^x \times N\times M}$ into a distance-matrix $\mathbf{Y}^{lk'}_{r}$ with the dimension of $M\times N_R^z N_R^xN$. Based on (45), $\mathbf{Y}^{lk'}_{r}$  can be represented as
\begin{equation}
	\begin{split}
		\begin{aligned}
			\label{deqn_ex1a}
			\mathbf{Y}^{lk'}_{r}= \mathbf{k}^{lk'}_{r}(r^{Long}_{k',l})\cdot
			\mathbf{x}^{lk'}_{r} + \mathbf{N}^{lk'}_{r} \in \mathbb{C}^{M\times N_R^z N_R^xN},
		\end{aligned}
	\end{split}\tag{52}
\end{equation}
where  $\mathbf{k}^{lk'}_{r}(r) = [1,e^{-j\frac{4\pi r\Delta f}{c}},...,e^{-j\frac{4\pi r\Delta f}{c}(M-1)}]^T \in \mathbb{C}^{M\times 1}$ is defined as the \emph{distance array steering vector}, $\mathbf{x}^{lk'}_{r}\in \mathbb{C}^{1\times N_R^z N_R^xN}$ and $\mathbf{N}^{lk'}_{r} \in \mathbb{C}^{M\times N_R^z N_R^xN}$.
Since $\mathbf{Y}^{lk'}_{r}$ is the array signals form related to the distance array, we can estimate $r^{Long}_{k',l}$ from $\mathbf{Y}^{lk'}_{r}$ by utilizing array signal processing methods.
Similarly, we can employ the ESPRIT method to obtain the space value  corresponding to $\mathbf{Y}^{lk'}_{r}$ as $\kappa_{r}^{lk'}$.
Then the polar distance of the $k'$-th dynamic target within the $l$-th TTS can be estimated as 
\begin{equation}
	\begin{split}
		\begin{aligned}
			\label{deqn_ex1a}
			\check{r}^{Long}_{k',l} =-\frac{c\kappa_{r}^{lk'}}{4\pi \Delta f}.
		\end{aligned}
	\end{split}\tag{53}
\end{equation}


\subsection{Radial Velocity  and Angular Velocities Estimation}

It can be analyzed from (45) that each antenna observes one \emph{virtual-velocity} composed of the radial velocity, horizontal angular velocity, and pitch angular velocity of the dynamic target. Note that the virtual-velocity  observed by different antenna is different.
We can design and derive the virtual-velocity of the $k'$-th dynamic target observed by the $(n^x_R,n^z_R)$-th antenna within the $l$-th TTS from (45) as $v^{Long,vir}_{k',l,n^x_R,n^z_R}$, which is shown in (54) at the top of this page.
Then (45) can be rewritten as (55) at the top of this page.

Next, we need to estimate the virtual-velocity observed by each antenna.
We can extract the DT-EEC of the $(n^x_R,n^z_R)$-th antenna on all subcarriers of all OFDM symbols from $\check{\mathbf{Y}}_{cube}^{lk'} \in \mathbb{C}^{N_{R}^z\times N_{R}^x \times N\times M}$ as
\begin{equation}
	\begin{split}
		\begin{aligned}
			\label{deqn_ex1a}
\mathbf{Y}^{lk'}_{v_{vir},n^x_R,n^z_R}\!=\! \mathbf{k}^{lk'}_{v_{vir}}(v^{Long,vir}_{k',l,n^x_R,n^z_R})\!\cdot\!
\mathbf{x}^{lk'}_{v_{vir},n^x_R,n^z_R} \!+\! \mathbf{N}^{lk'}_{v_{vir},n^x_R,n^z_R},
		\end{aligned}
	\end{split}\tag{56}
\end{equation}
where $\mathbf{Y}^{lk'}_{v_{vir},n^x_R,n^z_R} \in \mathbb{C}^{N\times M}$, $\mathbf{x}^{lk'}_{v_{vir}}\in \mathbb{C}^{1\times M}$, $\mathbf{N}^{lk'}_{v_{vir}} \in \mathbb{C}^{N\times M}$,  $[\mathbf{Y}^{lk'}_{v_{vir},n^x_R,n^z_R}]_{n,m}=\check{y}^{lk'}_{n,m,n^z_R,n^x_R}$, and 
$\mathbf{k}^{lk'}_{v_{vir}}(v_{vir}) = [1,e^{j\frac{4\pi f_0v_{vir}T_s}{c}},...,e^{j\frac{4\pi f_0v_{vir}T_s}{c}(N-1)}]^T \in \mathbb{C}^{N\times 1}$ is defined as the \emph{virtual-velocity array steering vector}. 
Since $\mathbf{Y}^{lk'}_{v_{vir},n^x_R,n^z_R}$ is the array signals form related to the virtual-velocity array, we can estimate $v^{Long,vir}_{k',l,n^x_R,n^z_R}$ from $\mathbf{Y}^{lk'}_{v_{vir},n^x_R,n^z_R}$ by utilizing array signal processing methods.
Similarly, we can employ the ESPRIT method to obtain the space value  corresponding to $\mathbf{Y}^{lk'}_{v_{vir},n^x_R,n^z_R}$ as $\kappa^{lk'}_{v_{vir},n^x_R,n^z_R}$.
Then the virtual-velocity of the $k'$-th dynamic target observed by the $(n^x_R,n^z_R)$-th antenna within the $l$-th TTS can be estimated as 
\begin{equation}
	\begin{split}
		\begin{aligned}
			\label{deqn_ex1a}
\check{v}^{Long,vir}_{k',l,n^x_R,n^z_R} =\frac{c \kappa^{lk'}_{v_{vir},n^x_R,n^z_R}}{4\pi f_0T_s}.
		\end{aligned}
	\end{split}\tag{57}
\end{equation}
By traversing each antenna, we can obtain the virtual-velocity observed by each antenna, 
which is record as 
$(n_R^x,n_R^z,\check{v}^{Long,vir}_{k',l,n^x_R,n^z_R})$ with  $n_{R}^x \in \{0,1,...,N^x_{R}-1\}$ and $n_{R}^z  \in \{0,1,...,N^z_{R}-1\}$.
Then we need to estimate the radial velocity, horizontal angular velocity, and pitch angular velocity of the target from these $N_R=N_R^xN_R^z$ ternary pairs.

In fact, we can express  $v^{Long,vir}_{k',l,n^x_R,n^z_R}$ as a binary function of $(n_R^x,n_R^z)$. Based on (54),
there is
\begin{equation}
	\begin{split}
		\begin{aligned}
			\label{deqn_ex1a}
{v}^{Long,vir}_{k',l,n^x_R,n^z_R} = A_{k',l}+B_{k',l}\cdot n^x_R+C_{k',l} \cdot n^z_R,
		\end{aligned}
	\end{split}\tag{58}
\end{equation}
where
\begin{equation}
	\begin{split}
		\begin{aligned}
			\label{deqn_ex1a}
A_{k'\!,l}\!=&  v^{L\!o\!n\!g}_{r\!,k'\!,l} \!\!-\!\!\frac{d}{4}\!\!\left[\!(\!N_H^z\!\!-\!\!1)\!\cos\!\phi^{L\!o\!n\!g}_{k'\!,l}
\!\!-\!\!(\!N_H^x\!\!-\!\!1\!)\!\sin\!\phi^{L\!o\!n\!g}_{k',l}\!\cos\!\theta^{L\!o\!n\!g}_{k',l}\!\right]\!\!\omega^{L\!o\!n\!g}_{\phi,k'\!,l}
\\& +\frac{d}{4}
(N^x_H\!-\!1)\cos\phi^{Long}_{k',l}\sin\theta^{Long}_{k',l}\omega^{Long}_{\theta,k',l},
		\end{aligned}
	\end{split}\tag{59}
\end{equation}
\begin{equation}
	\begin{split}
		\begin{aligned}
			\label{deqn_ex1a}
B_{k'\!,l}\!=&  \frac{d}{2} (\sin\!\phi^{L\!o\!n\!g}_{k',l}\cos\!\theta^{L\!o\!n\!g}_{k',l}\omega^{L\!o\!n\!g}_{\phi,k'\!,l}\!+\!
\cos\phi^{Long}_{k',l}\sin\theta^{Long}_{k',l}\omega^{Long}_{\theta,k',l}),
		\end{aligned}
	\end{split}\tag{60}
\end{equation}
\begin{equation}
	\begin{split}
		\begin{aligned}
			\label{deqn_ex1a}
\!\!\!\!\!\!\!\!\!\!\!\!\!\!\!\!\!\!\!\!
\!\!\!\!\!\!\!\!\!\!\!\!\!\!\!\!\!\!\!\!
\!\!\!\!\!\!\!\!\!\!\!\!\!\!\!\!\!\!\!\!
\!\!\!\!\!\!\!\!\!\!\!\!\!\!\!
C_{k',l}\!=&  -\frac{d}{2} \cos\phi^{Long}_{k',l} \omega^{Long}_{\phi,k'\!,l}.
		\end{aligned}
	\end{split}\tag{61}
\end{equation}
Formula (58) indicates that  ternary pairs $(n_R^x,n_R^z,{v}^{Long,vir}_{k',l,n^x_R,n^z_R})$ could form a plane in three-dimensional space.

Therefore, we can   use the least squares (LS) method for planar fitting of $\{(n_R^x,n_R^z,\check{v}^{Long,vir}_{k',l,n^x_R,n^z_R})|n_{R}^x =0,1,...,N^x_{R}-1; n_{R}^z =0,1,...,N^z_{R}-1\}$, and  we record the parameter results of plane fitting as
$\check{A}_{k',l}$, $\check{B}_{k',l}$ and
$\check{C}_{k',l}$. 
Then based on (59), (60) and (61),
the pitch angular velocity, horizontal angular velocity, and radial velocity of the $k'$-th dynamic target within the $l$-th TTS can be sequentially estimated as
\begin{equation}
	\begin{split}
		\begin{aligned}
			\label{deqn_ex1a}
\!\!\!\!\!\!\!\!\!\!\!\!\!\!\!\!\!\!\!\!\!\!\!\!\!\!\!\!\!\!\!\!\!\!\!\!\!\!\!\!\!\!\!\!\!\!\!\!\!\!\!\!\!\!\!\!\!\!\!\!\!\!\!\!\!\!\!\!\check{\omega}_{\phi,k',l}^{Long}\!=&  -\frac{2\check{C}_{k',l}}{d\cos\phi^{Long}_{k',l}},
		\end{aligned}
	\end{split}\tag{62}
\end{equation}
\begin{equation}
	\begin{split}
		\begin{aligned}
			\label{deqn_ex1a}
\check{\omega}_{\theta,k',l}^{Long}\!=&  -\frac{2\check{B}_{k',l}/d -\sin\check{\phi}^{Long}_{k',l}\cos\check{\theta}^{Long}_{k',l} \check{\omega}_{\phi,k',l}^{Long}}{\cos\check{\phi}^{Long}_{k',l}\sin\check{\theta}^{Long}_{k',l}},
		\end{aligned}
	\end{split}\tag{63}
\end{equation}
\begin{equation}
	\begin{split}
		\begin{aligned}
			\label{deqn_ex1a}
\!\!\!\!&\check{v}_{r,k',l}^{Long}\!\!=\!\! \check{A}_{k'\!,l}\!+\!
\!\frac{d}{4}\!\!\left[\!(\!N_{\!H}^z\!\!-\!\!1\!)\!\cos\!\check{\phi}^{L\!o\!n\!g}_{k'\!,l}
\!\!-\!\!(\!N_{\!H}^x\!-\!\!1\!)\!\sin\!\check{\phi}^{L\!o\!n\!g}_{k',l}\!\!\cos\!\check{\theta}^{L\!o\!n\!g}_{k',l}\!\right]\!\!\check{\omega}^{L\!o\!n\!g}_{\phi,k'\!,l} \\&         	
\kern 35pt -\frac{d}{4}
(N^x_H\!-\!1)\cos\check{\phi}^{Long}_{k',l}\sin\check{\theta}^{Long}_{k',l}\check{\omega}^{Long}_{\theta,k',l}.
		\end{aligned}
	\end{split}\tag{64}
\end{equation}

\begin{figure}[!t]
	\centering
	\includegraphics[width=80mm]{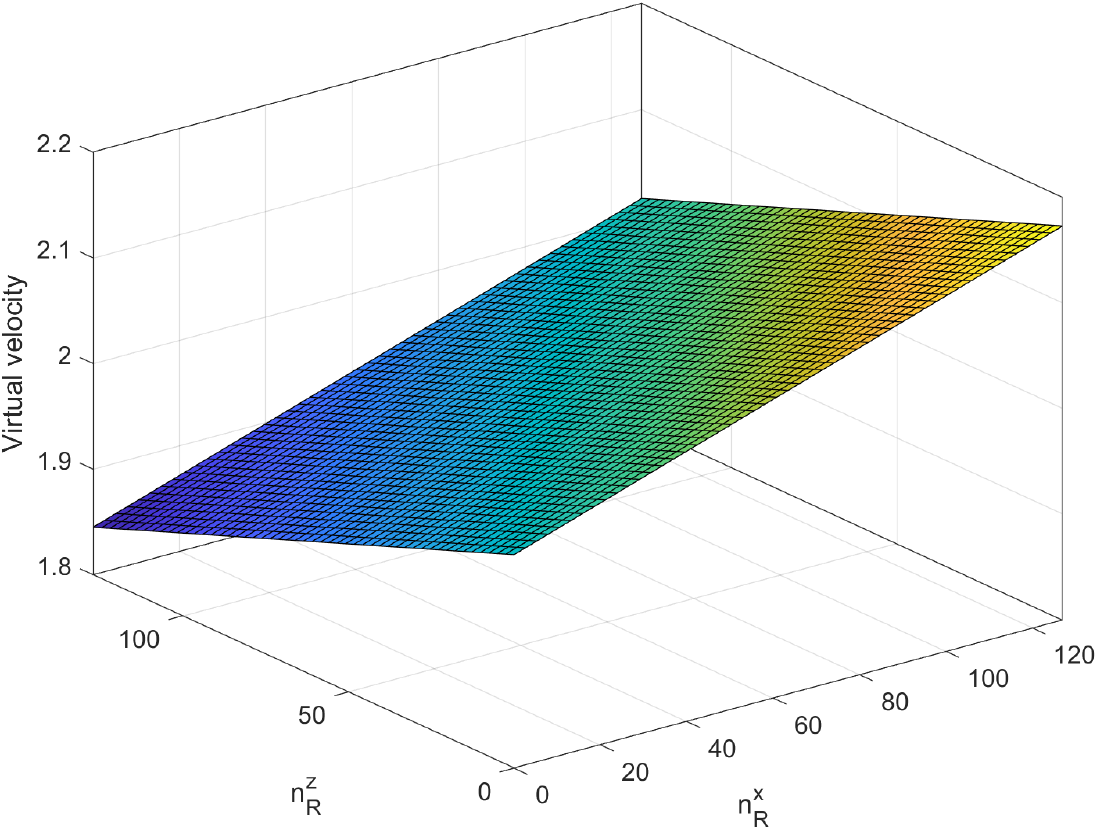}
	\caption{Virtual velocity observed by different antennas.}
	\label{fig_1}
\end{figure}

Based on (48), (51), (53), (62), (63) and (64),
we have estimated the 6D motion parameters of the $k'$-th dynamic target within the $l$-th TTS as
$\check{\mathbf{S}}^{Long}_{k',l} = [\check{r}^{Long}_{k',l},\check{\theta}_{k',l}^{Long},\check{\phi}_{k',l}^{Long},\check{v}_{r,k',l}^{Long},\check{\omega}_{\theta,k',l}^{Long},\check{\omega}_{\phi,k',l}^{Long}]^T$.

Fig.~4 shows an example of virtual velocity estimation and fitting.  It can be seen from the figure that different antennas have observed different virtual velocities for the same dynamic target,  the 2D antenna index and virtual velocity form a plane in the 3D coordinate system, and thus we
can   recover the radial velocity, horizontal angular velocity, and pitch angular velocity of the target from these virtual velocities $\{(n_R^x,n_R^z,\check{v}^{Long,vir}_{k',l,n^x_R,n^z_R})|n_{R}^x =0,1,...,N^x_{R}-1; n_{R}^z =0,1,...,N^z_{R}-1\}$ based on formulas from (58) to (64).

\subsection{Long-Term Motion Tracking}

We consider the 6D motion parameters of the $k$-th dynamic target $\mathbf{S}^{Long}_{k,l}$ as the state of one micro-system, and consider the $\check{\mathbf{S}}^{Long}_{k,l}$  obtained through the 6D  sensing  algorithm as the observation of this micro-system. Based on the formulas from
(8) to (13), the state equation  can be expressed as
\begin{equation}
	\begin{split}
		\begin{aligned}
			\label{deqn_ex1a}
\mathbf{S}^{Long}_{k,l+1}=\mathbf{\Phi}
\mathbf{S}^{Long}_{k,l}+
\mathbf{B}\mathbf{u}^{Long}_{k,l}+
\mathbf{w}_{k,l}^{Long},
		\end{aligned}
	\end{split}\tag{65}
\end{equation}
where $\mathbf{\Phi} = 
\begin{bmatrix} 
	\mathbf{I}_{3\times 3}& -T_{TTS}\mathbf{I}_{3\times 3}
	\\\mathbf{0}_{3\times 3}&\mathbf{I}_{3\times 3} 
\end{bmatrix}$
is  state transition matrix,
$\mathbf{B} = 
\begin{bmatrix} 
	-\frac{1}{2}T_{TTS}^2\mathbf{I}_{3\times 3}
	\\-T_{TTS}\mathbf{I}_{3\times 3}
\end{bmatrix}$
is  disturbance driven matrix,
$\mathbf{w}_{k,l}^{Long}$ is the state noise matrix and its  covariance matrix is $\mathbf{Q}$.
Besides, the observation equation of the micro-system can be represented as
\begin{equation}
	\begin{split}
		\begin{aligned}
			\label{deqn_ex1a}
\check{\mathbf{S}}^{Long}_{k,l}=\mathbf{G}
\mathbf{S}^{Long}_{k,l}+
			\mathbf{v}_{k,l}^{Long},
		\end{aligned}
	\end{split}\tag{66}
\end{equation}
where $\mathbf{G}=\mathbf{I}_{6\times 6}$ is the observation matrix,
$\mathbf{v}_{k,l}^{Long}$ is equivalent observation noise vector and its covariance matrix is $\mathbf{R}$.
Then we can use  Kalman filtering (KF) \cite{Ma2020} to track the long-term motion of the $k$-th dynamic target as follows:

1) \emph{Initialization}:
ISAC BS can obtain the 6D parameters estimation $\mathbf{S}^{SBS}_{k}$ of the $k$-th dynamic target through beam  scanning during  SBS stage.
Next, to enter the SBT stage,
we initialize  the time as $l = 0$,
 the observation as 
$\check{\mathbf{S}}^{Long}_{k,0} = \mathbf{S}^{SBS}_{k}$,
 the state estimation as
$\hat{\mathbf{S}}^{Long}_{k,0} = \mathbf{S}^{SBS}_{k}$,
and  $\hat{\mathbf{P}}_{k,0}=\mathbf{I}_{6\times 6}$.

2) \emph{State prediction}: 
Based on  $\hat{\mathbf{S}}^{Long}_{k,l-1}$,  the state prediction  within  the $l$-th TTS can be calculated as 
$\tilde{\mathbf{S}}^{Long}_{k,l}=\mathbf{\Phi}\hat{\mathbf{S}}^{Long}_{k,l-1}$.

3) \emph{Observation prediction}: The observation prediction within the $l$-th TTS can be computed
as $\tilde{\check{\mathbf{S}}}^{Long}_{k,l}=\mathbf{G}\tilde{\mathbf{S}}^{Long}_{k,l}$.

4) \emph{Calculate Kalman gain}:
Based on $\hat{\mathbf{P}}_{k,l-1}$, we can compute $\tilde{{\mathbf{P}}}_{k,l}=\mathbf{\Phi}\hat{\mathbf{P}}_{k,l-1}\mathbf{\Phi}^T$. Then the Kalman gain can be obtained as $\mathbf{K}^{gain}_{k,l}=\tilde{{\mathbf{P}}}_{k,l}\mathbf{G}^T
\left(\mathbf{G}\tilde{{\mathbf{P}}}_{k,l}\mathbf{G}^T+\mathbf{R} \right)^{-1}$.

5) \emph{State estimation update}:
The KF estimation of the 6D motion  parameters   can be updated and  represented as
$\hat{\mathbf{S}}^{Long}_{k,l} =\tilde{\mathbf{S}}^{Long}_{k,l}+
\mathbf{K}^{gain}_{k,l}({\check{\mathbf{S}}}^{Long}_{k,l}-\tilde{\check{\mathbf{S}}}^{Long}_{k,l})$. 
Besides, $\hat{\mathbf{P}}_{k,l}$ can be updated as $\hat{\mathbf{P}}_{k,l}=(\mathbf{I}_{6\times 6}-\mathbf{K}^{gain}_{k,l}\mathbf{G})\tilde{{\mathbf{P}}}_{k,l}$.

Based on the above  steps, we can continuously track the dynamic targets within $L$ TTS, and we employ $\hat{\mathbf{S}}^{Long}_{k,l}$ as the final 6D motion parameters estimation result of the $k$-th dynamic target within the $l$-th TTS.

\begin{figure*}[!t]
	\centering
	\subfloat[]{\includegraphics[width=65mm]{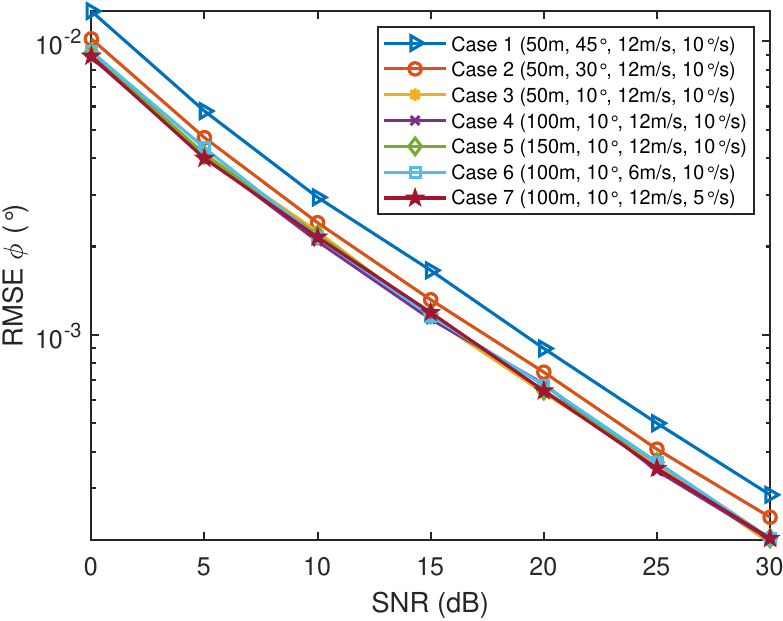}%
		\label{fig_first_case}}
	\hfil
	\subfloat[]{\includegraphics[width=65mm]{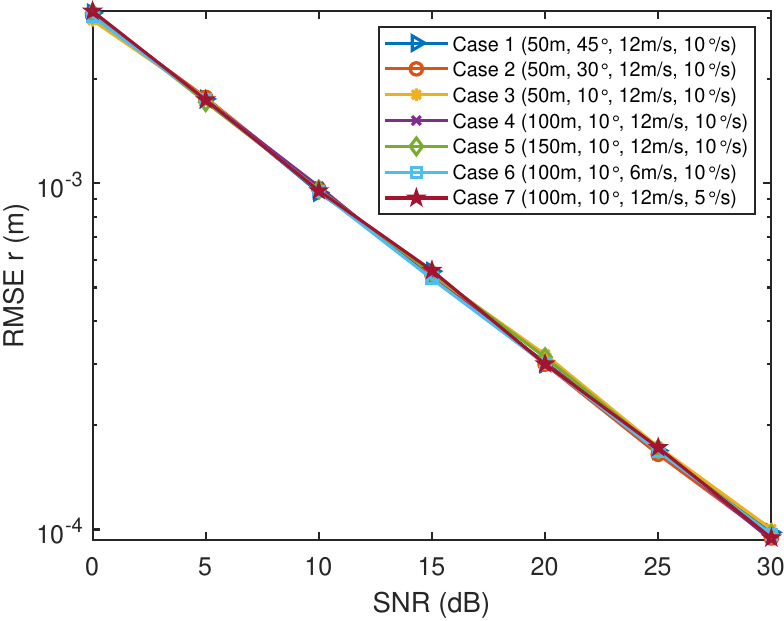}%
		\label{fig_first_case}}
	\hfil
	\subfloat[]{\includegraphics[width=65mm]{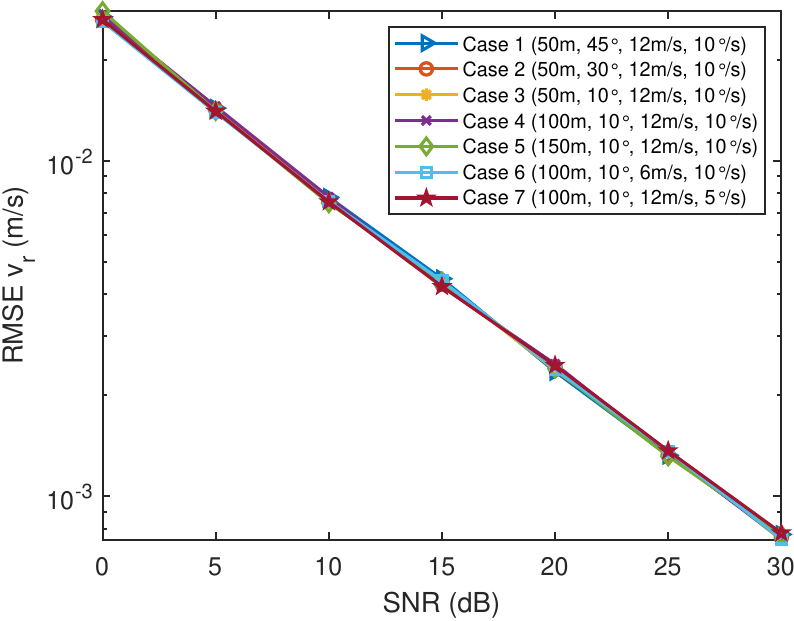}%
		\label{fig_first_case}}
	\hfil
	\subfloat[]{\includegraphics[width=65mm]{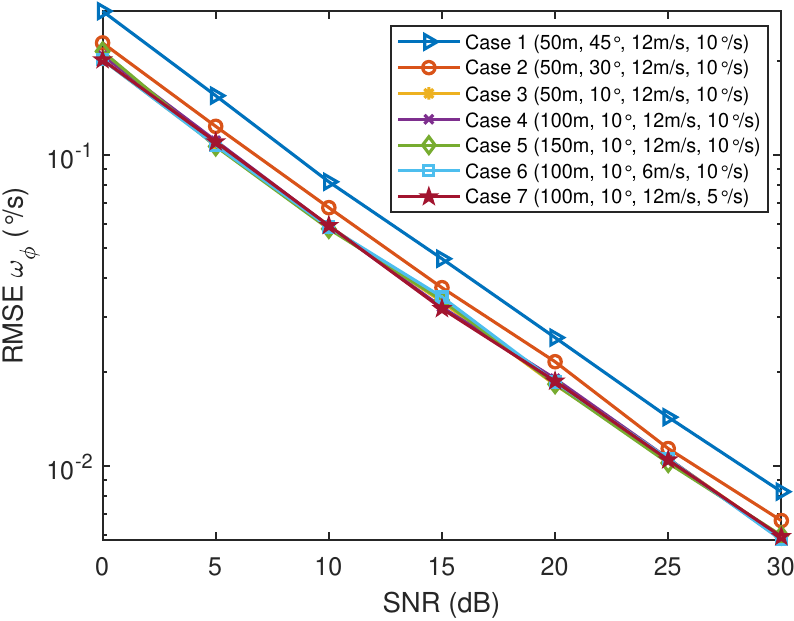}%
		\label{fig_second_case}}
	\caption{(a) The ${\rm RMSE}_\phi$ versus SNR curves of different targets. (b) The ${\rm RMSE}_r$ versus SNR curves of different targets.
		(c) The ${\rm RMSE}_{v_r}$ versus SNR curves of different targets.
		(d) The ${\rm RMSE}_{\omega_\phi}$ versus SNR curves of different targets.}
	\label{fig_sim}
\end{figure*}

\section{Simulation Results}

In simulations,
 we set 
the lowest carrier frequency of the ISAC system as $f_0 = 100$ GHz,
set the subcarrier frequency interval as $\Delta f = 480$ kHz,
 and set the antenna spacing as $d=\frac{1}{2}\lambda$.
To succinctly display the simulation results, we set the horizontal angle and horizontal angular velocity of the dynamic target as
$\theta_k = 90^\circ$ and $\omega_{\theta,k}=0$,
and thus the dynamic target is fixed to move within a 2D plane. Then we  focus on the sensing accuracy of the $\{r_k,  \phi_k, v_{r,k},  \omega_{\phi,k}\}$ parameters of the dynamic target.

Specifically,  for the aspect of evaluating  6D radar sensing,
the root  mean square error (RMSE) of
distance sensing,  angle sensing,  radial velocity sensing and  angular velocity sensing
 are defined as
 ${\rm RMSE}_r=\sqrt{\frac{\sum _{i=1}^{Count}(\check{r}_{s(i)}-r_{s})^2}{Count}}$,
${\rm RMSE}_\phi=\sqrt{\frac{\sum _{i=1}^{Count}(\check{\phi}_{s(i)}-\phi_{s})^2}{Count}}$,   ${\rm RMSE}_{v_r}=\sqrt{\frac{\sum _{i=1}^{Count}(\check{v}_{r,s(i)}-v_{r,s})^2}{Count}}$, and ${\rm RMSE}_{\omega_\phi}=\sqrt{\frac{\sum _{i=1}^{Count}(\check{\omega}_{\phi,s(i)}-\omega_{\phi,s})^2}{Count}}$, 
where $Count$ is the number of the Monte Carlo runs,
the real parameters of the dynamic target is
$(r_{s},\phi_{s},v_{r,s},\omega_{\phi,s})$, and $(\check{r}_{s},\check{\phi}_{s},\check{v}_{r,s},\check{\omega}_{\phi,s})$ is the estimation parameters of the target.

\begin{figure*}[!t]
\centering
\subfloat[]{\includegraphics[width=60mm]{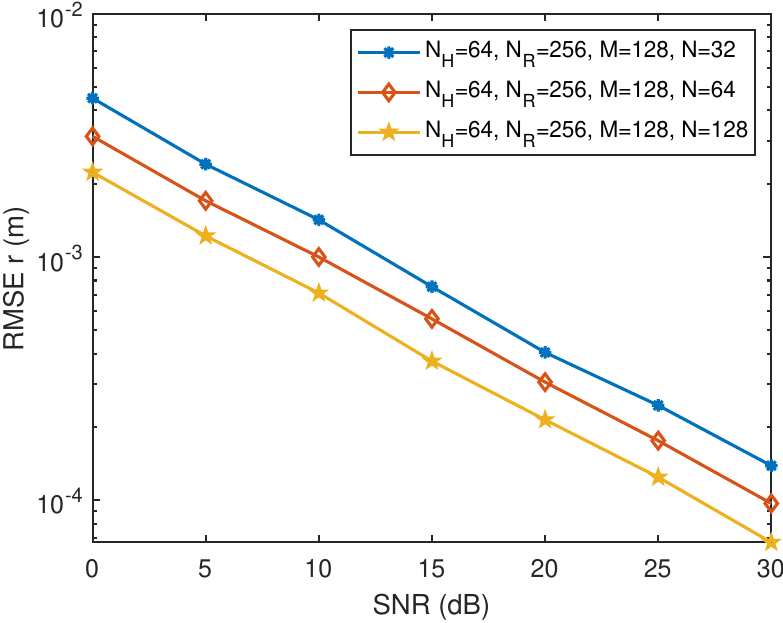}%
\label{fig_first_case}}
\hfil
\subfloat[]{\includegraphics[width=60mm]{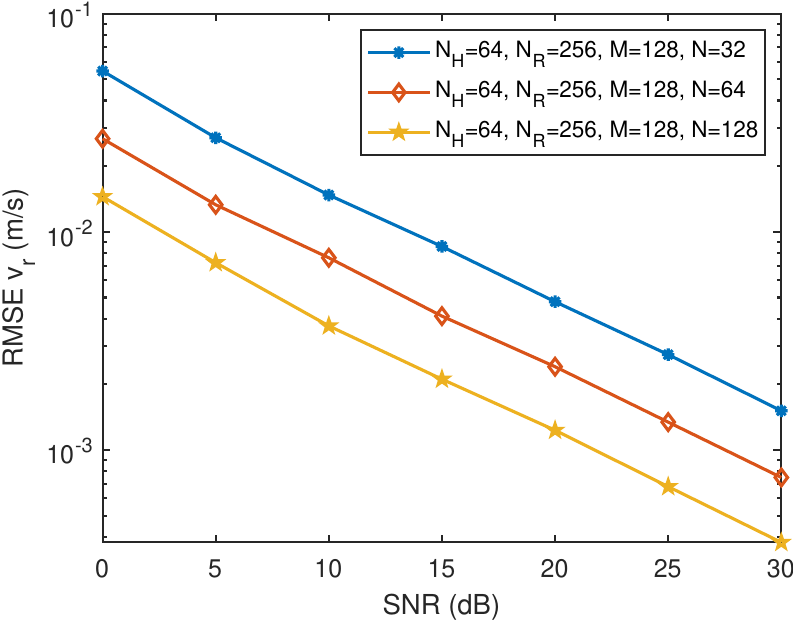}%
\label{fig_first_case}}
\hfil
\subfloat[]{\includegraphics[width=60mm]{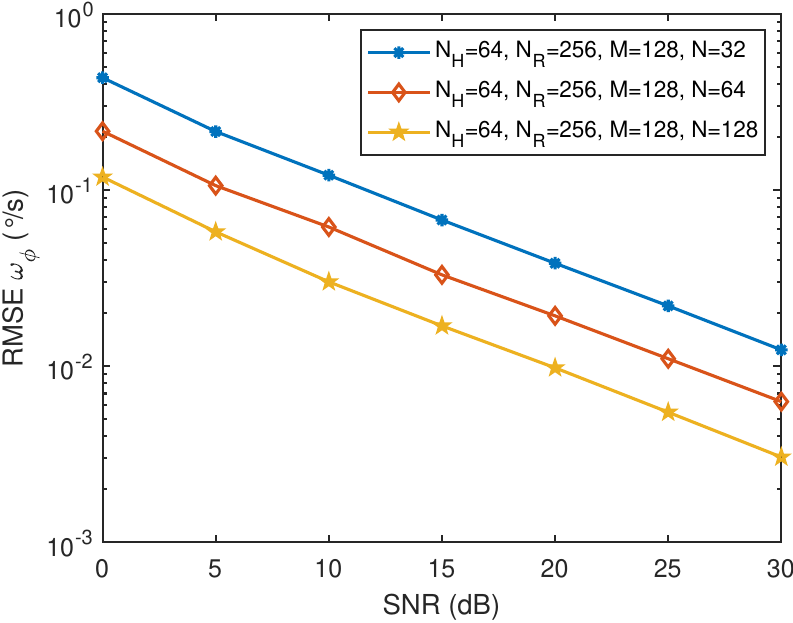}%
\label{fig_second_case}}
\caption{(a) Distance sensing performance under different numbers of OFDM symbols.
(b) Radial velocity sensing performance under different numbers of OFDM symbols.
(c) Angular  velocity sensing performance under different numbers of OFDM symbols.}
\label{fig_sim}
\end{figure*}

\begin{figure*}[!t]
\centering
\subfloat[]{\includegraphics[width=60mm]{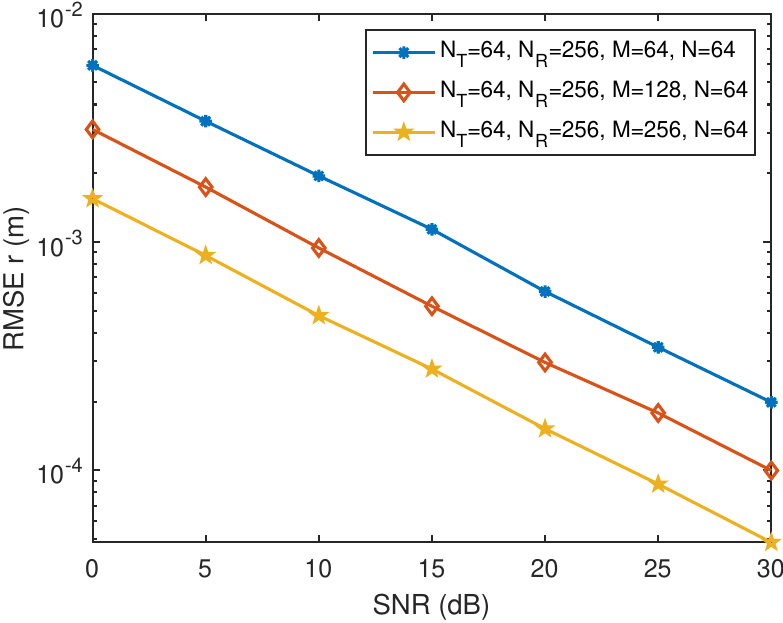}%
\label{fig_first_case}}
\hfil
\subfloat[]{\includegraphics[width=60mm]{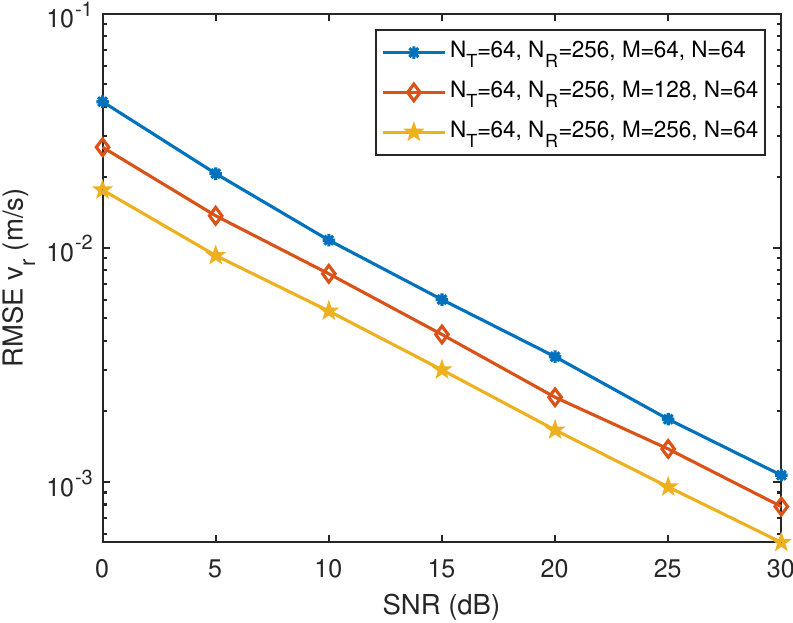}%
\label{fig_first_case}}
\hfil
\subfloat[]{\includegraphics[width=60mm]{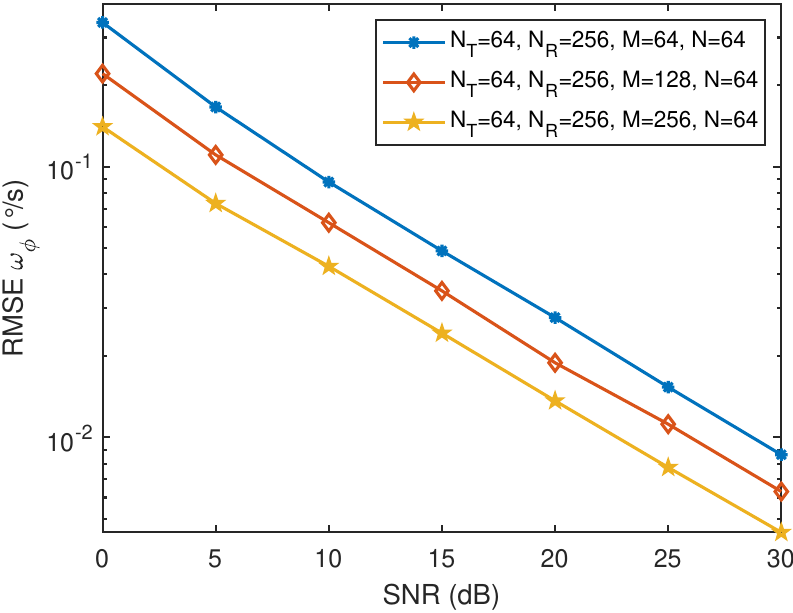}%
\label{fig_second_case}}
\caption{(a) Distance sensing performance under different numbers of subcarriers.
	(b) Radial velocity sensing performance under different numbers of subcarriers.
	(c) Angular  velocity sensing performance under different numbers of subcarriers.}
\label{fig_sim}
\end{figure*}

\subsection{The Performance of 6D Radar Single-Shot Sensing}

We set that 
the number of subcarriers is $M=128$,
 the number of OFDM symbols is $N=64$,
the number of the antennas in HU-UPA is $N_H=64$,
the number of the antennas in RU-UPA is $N_R=256$.
Fig.~5 shows the single-shot sensing RMSE of the proposed scheme for
different dynamic targets  with different motion parameters versus SNR.
It can be seen  that the ${\rm RMSE}_{\phi}$, ${\rm RMSE}_{r}$, ${\rm RMSE}_{v_r}$, and ${\rm RMSE}_{\omega_{\phi}}$ gradually decrease with the increase of SNR. When SNR $=0$ dB, the average sensing RMSEs are
${\rm RMSE}_{\phi}=0.0097^\circ$, ${\rm RMSE}_{r}=0.0031m$, ${\rm RMSE}_{v_r}=0.0267m/s$, and ${\rm RMSE}_{\omega_{\phi}}=0.2208^\circ/s$.
When SNR increases to $20$ dB, the average sensing RMSEs decrease to
${\rm RMSE}_{\phi}=0.0007^\circ$, ${\rm RMSE}_{r}=0.0003m$, ${\rm RMSE}_{v_r}=0.0024m/s$, and ${\rm RMSE}_{\omega_{\phi}}=0.0200^\circ/s$.

Unlike most existing ISAC studies  believing that only the radial velocity of far-field dynamic target can be measured based on one single BS. 
These simulation results indicate that the proposed 6D radar sensing algorithm has high sensing accuracy, especially confirming that one single BS with MIMO array can effectively estimate the angular velocity of the dynamic target.

Besides,
 it is found from Fig.~5(b) and  Fig.~5(c) that under the same system parameter settings, the distance sensing and radial velocity sensing performance of dynamic targets with different motion parameters are basically consistent. However, it is seen from Fig.~5(a) and  Fig.~5(d) that
 under the same system parameter settings,  the accuracy of dynamic target angle sensing and angular velocity senisng gradually improves as the target approaches $0^\circ$, mainly because the MIMO array has  narrower beamwidth near $0^\circ$, thus improving the accuracy of angle sensing. Since the angular velocity sensing depends on the angle change of the target, the  narrower beam near $0^\circ$ also brings higher angular velocity sensing accuracy.

\subsection{The Impact of System Parameters on the Performance of 6D Radar Single-Shot Sensing}

We take the sensing of  the  dynamic target with motion parameters ($\phi=20^\circ$,
$r=120m$, $v_r=15m/s$, $\omega_{\phi}=8^\circ/s$) as the example, and
  investigate the impact of system parameter settings on the performance of 6D radar single-shot sensing.

Fig.~6 shows the variation curves of distance sensing, radial velocity sensing, and angular velocity sensing versus SNR under different number of  OFDM symbols. It can be seen from Fig.~6(a)  that the ${\rm RMSE}_r$ gradually decreases as the number of OFDM symbols $N$ increases. This is because more OFDM symbols bring more observations to the distance array, making the estimation of the covariance matrix of the distance array more accurate, and  thereby improving the accuracy of distance sensing.
Besides,  it can be found from Fig.~6(b) and Fig.~6(c) that the ${\rm RMSE}_{v_r}$ and the ${\rm RMSE}_{\omega_\phi}$ significantly decrease with the increase of $N$. This is because more OFDM symbols form a larger virtual velocity array, making the sensing of radial velocity and angular velocity more accurate.

\begin{figure*}[!t]
\centering
\subfloat[]{\includegraphics[width=60mm]{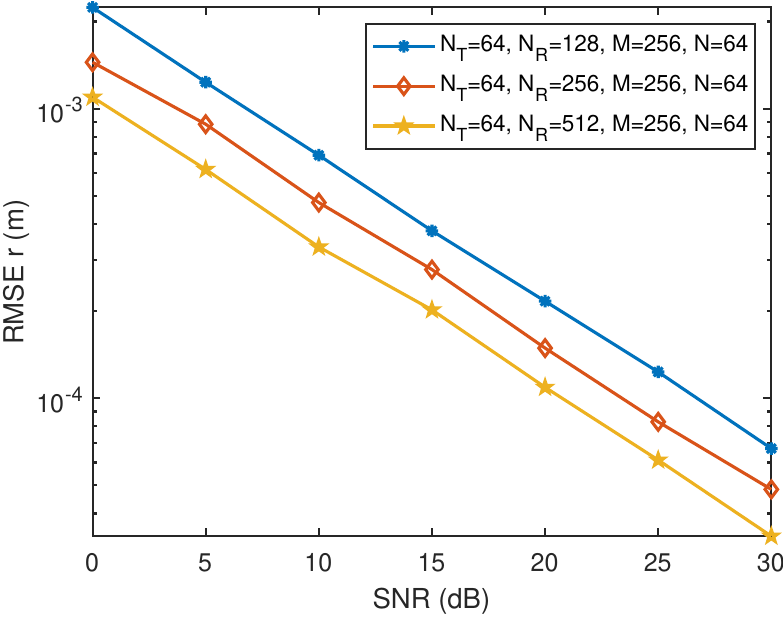}%
\label{fig_first_case}}
\hfil
\subfloat[]{\includegraphics[width=60mm]{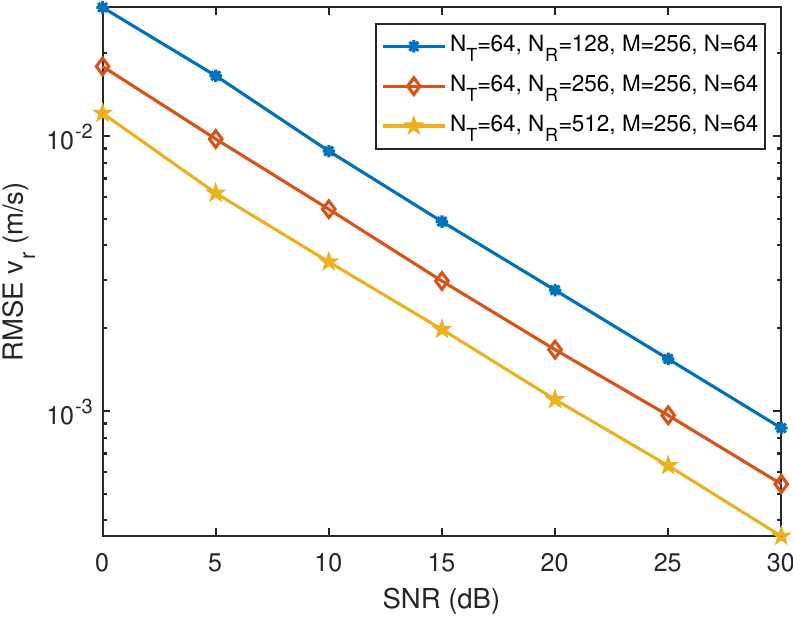}%
\label{fig_first_case}}
\hfil
\subfloat[]{\includegraphics[width=60mm]{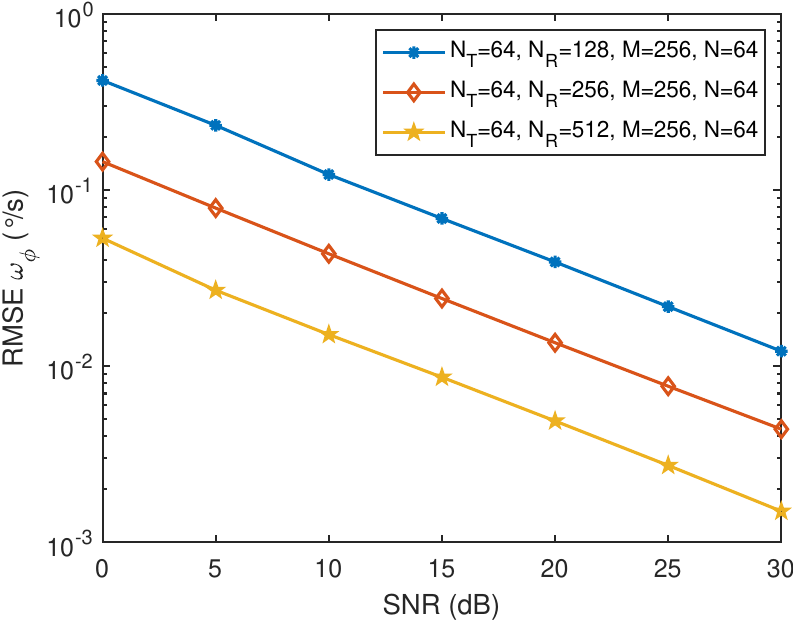}%
\label{fig_second_case}}
\caption{(a) Distance sensing performance under different numbers of receiving antennas.
	(b) Radial velocity sensing performance under different numbers of receiving antennas.
	(c) Angular  velocity sensing performance under different numbers of receiving antennas.}
\label{fig_sim}
\end{figure*}

\begin{figure*}[!t]
	\centering
	\includegraphics[width=170mm]{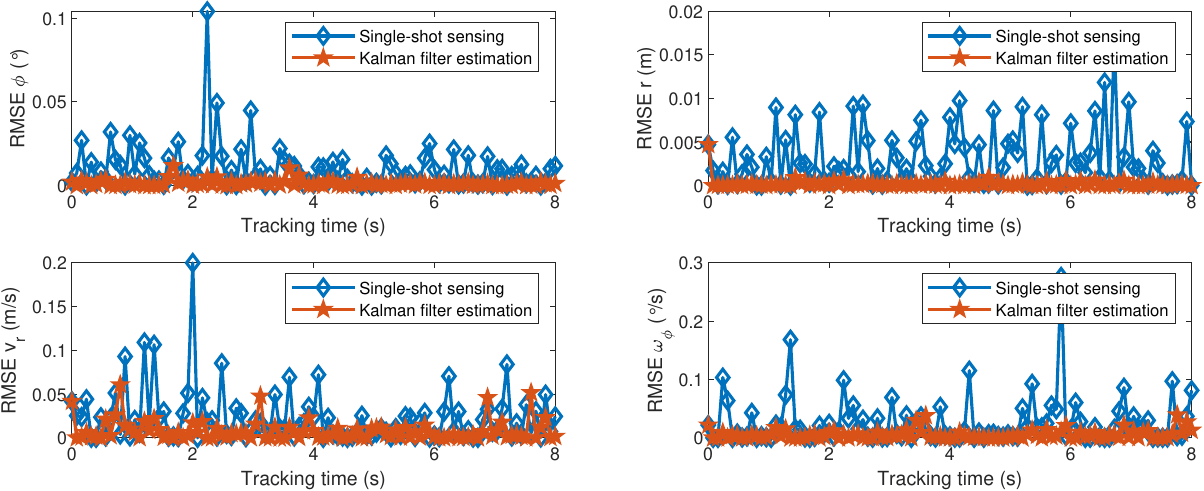}
	\caption{The performance of multiple-shots tracking.}
	\label{fig_1}
\end{figure*}

Fig.~7 shows the variation curves of  sensing RMSEs versus SNR under different number of  subcarriers. 
 It can be seen from Fig.~7(a) that the ${\rm RMSE}_r$ gradually decreases with the increase of the number of subcarriers $M$, because more subcarriers can form a larger distance array, thereby improving the accuracy of distance sensing. It can be found from  Fig.~7(b) and Fig.~7(c) that the  ${\rm RMSE}_{v_r}$ and the ${\rm RMSE}_{\omega_\phi}$  gradually decrease with the increase of $M$. This is because more subcarriers bring more observations to the virtual velocity array, making the covariance matrix estimation of the virtual velocity array more accurate, thereby improving the sensing accuracy of radial velocity and angular velocity.

Fig.~8 shows the variation curves of  sensing RMSEs versus SNR under different number of antennas.
 It can be seen from Fig.~8(a) that ${\rm RMSE}_r$ gradually decreases as the number of antennas $N_R$ increases. This is because more receiving antennas provide more observations for  distance array, thereby improving the accuracy of distance sensing.
More importantly, it can be observed from  Fig.~8(b) and Fig.~8(c) that the  ${\rm RMSE}_{v_r}$ and the ${\rm RMSE}_{\omega_\phi}$ gradually decrease with the increase of $N_R$.
This is because when there are more receiving antennas measuring the virtual velocity, the system can better fit the virtual velocity plane, thereby more accurately recovering the radial velocity and angular velocity of dynamic target.

\subsection{The Performance of Multiple-Shots Tracking}

We  set the dynamic target with the initial motion parameter of $\phi=55^\circ$,
$r=100m$, $v_r=8m/s$, $\omega_{\phi}=4^\circ/s$, and the BS needs to track this target within $8$ seconds.
We set the system parameters as
$M=128$, $N=64$, $N_H=64$, and $N_R=256$.
Fig.~9 shows the tracking performance of multiple-shots tracking for dynamic target when SNR $=0$ dB.
It can be seen from the figure  that the dynamic target tracking algorithm based on Kalman filtering can further improve the sensing accuracy of 6D radar single-shot sensing, especially the performance of angular velocity sensing. These simulation results verify the effectiveness of the proposed scheme.

\section{Conclusions}

In this paper, 
we have proposed a novel scheme for 6D radar sensing and tracking of dynamic target based on MIMO array for monostatic ISAC system.
We have re-examined and re-derived the relationship between  6D motion parameters of dynamic target  and  sensing echo channel of MIMO-ISAC system, and
 found that the sensing echo channel  actually includes the distance, horizontal angle, pitch angle, radial velocity, horizontal angular velocity, and pitch angular velocity of  dynamic target. 
Specifically, we have proposed the 6D long-term motion and short-term motion model of dynamic target.
Then we have derived the   sensing channel model corresponding to  short-term motion. 
Next, for single-shot sensing, we employed the  array signal processing methods to estimate the dynamic target's distance, horizontal angle, pitch angle, and virtual velocity. Then we found that the virtual velocities observed by different antennas were different, which allowed us to utilize plane parameter fitting to estimate the radial velocity, horizontal angular velocity, and pitch angular velocity of the dynamic target.
Furthermore, we have realized the multiple-shots tracking of dynamic target  based on   each single-shot sensing results and Kalman filtering.
Simulation results have been provided to demonstrate the effectiveness of the proposed 6D radar sensing and tracking scheme.

\bibliographystyle{ieeetr}
\bibliography{paper9.bib}

\vfill

\end{document}